\documentclass[american,british,french,english,review, authoryear]{elsarticle}
\usepackage[T1]{fontenc}
\usepackage[latin9]{inputenc}
\usepackage{babel}
\addto\extrasfrench{%
   \providecommand{\fg}{\ifdim\lastskip>\z@\unskip\fi~\frqq}%
}

\usepackage{units}
\usepackage{textcomp}
\usepackage{url}
\usepackage{amsmath}
\usepackage{amssymb}
\usepackage{graphicx}
\PassOptionsToPackage{normalem}{ulem}
\usepackage{ulem}
\usepackage{subscript}
\usepackage[unicode=true,
 bookmarks=true,bookmarksnumbered=false,bookmarksopen=false,
 breaklinks=false,pdfborder={0 0 1},backref=false,colorlinks=false]
 {hyperref}
\hypersetup{pdftitle={Retrieving Atmospheric Dust Opacity on Mars by Imaging Spectroscopy at Large Angles},
 pdfauthor={Douté et al.},
 pdfkeywords={Mars atmosphere imaging spectrocopy aerosols OMEGA CRISM}}

\makeatletter

\providecommand{\tabularnewline}{\\}

\newcommand\micron{\mbox{$\mu$m}}%
\newcommand{\co}{CO\ensuremath{_{2}}}%
\newcommand{\ho}{H\ensuremath{_{2}}O}%

\makeatother

\begin{document}

\title{Retrieving Atmospheric Dust Opacity on Mars by Imaging Spectroscopy
at Large Angles}

\author{S. Douté$^{1}$}

\address{$^{1}$Institut de Planétologie et d'Astrophysique de Grenoble (IPAG),
France (sylvain.doute@obs.ujf-grenoble.fr Phone: +33 4 76 51 41 71
Fax +33 4 76 51 41 46)}

\author{X. Ceamanos$^{2}$}

\address{$^{2}$Meteo France CNRM/GMME/VEGEO}

\author{T. Apperé$^{3}$}

\address{$^{3}$Laboratoire AIM CEA-Saclay}
\begin{abstract}
  We propose a new method to retrieve the optical depth of Martian aerosols (AOD) from OMEGA and CRISM hyperspectral imagery at a reference
  wavelength of 1 \micron. Our method works even if the underlying surface is completely made of minerals, corresponding to a low contrast
  between surface and atmospheric dust, while being observed at a fixed geometry. Minimizing the effect of the surface reflectance
  properties on the AOD retrieval is the second principal asset of our method. The method is based on the parametrization of the radiative
  coupling between particles and gas determining, with local altimetry, acquisition geometry, and the meteorological situation, the
  absorption band depth of gaseous \co.  Because the last three factors can be predicted to some extent, we can define a new parameter
  $\beta$ that expresses specifically the strength of the gas-aerosols coupling while directly depending on the AOD. Combining estimations
  of $\beta$ and top of the atmosphere radiance values extracted from the observed spectra within the \co\ gas band at 2 \micron, we
  evaluate the AOD and the surface reflectance by radiative transfer inversion. One should note that practically $\beta$ can be estimated
  for a large variety of mineral or icy surfaces with the exception of CO\textsubscript{2} ice when its 2 \micron\ solid band is not
  sufficiently saturated. Validation of the proposed method shows that it is reliable if two conditions are fulfilled: (i) the observation
  conditions provide large incidence or/and emergence angles (ii) the aerosol are vertically well mixed in the atmosphere.  Experiments
  conducted on OMEGA nadir looking observations as well as CRISM multi-angular acquisitions with incidence angles higher than
  65\textdegree{} in the first case and 33\textdegree{} in the second case produce very satisfactory results. Finally in a companion paper
  the method is applied to monitoring atmospheric dust spring activity at high southern latitudes on Mars using OMEGA.\end{abstract}
\maketitle

\section*{Introduction}

Visible and near infrared imaging spectroscopy is a key remote sensing
technique to study and monitor the planet Mars. Although its atmosphere
is much fainter than Earth's, its composition dominated by \co\ gas
implies numerous and strong absorption bands that often overlap with
spectral features coming from the surface. Furthermore small mineral
particles or \ho\ ice clouds often drift over Martian surfaces at
various altitudes. These aerosols have also a strong, spatially and
temporally varying influence on the morphology of the acquired spectra.
As a consequence accurate analysis for the study of surface materials
requires the modeling and the correction of the atmospheric spectral
effects. The first step in this matter consists in retrieving the
aerosol optical depth (AOD) over the scene.

Since 2004 the imaging spectrometer OMEGA aboard Mars Express performs
nadir-looking and EPF (Emission Phase Function) observations in the
VIS (visible) and the SWIR (short wave infrared) (920-5100 nm at 14
to 23 nm spectral resolution) for the study of the surface and the
atmosphere of the red planet. See \citet{Clancy1991} for the definition
of the EPF mode. The spatial resolution of OMEGA typically varies
between 300 and 2000 m/pixel due to its eccentric orbit. The authors
in \citet{Vincendon2007a} have developed a method to quantify the
contribution of atmospheric dust in SWIR spectra obtained by OMEGA
regardless of the Martian surface composition. Using multi-temporal
observations at nadir with significant differences in solar incidence
angles, they can infer the AOD and retrieve the surface reflectance
spectra free of aerosol contribution. However, this method relies
on the very restrictive assumption that the atmosphere opacity remains
approximately constant during the time spanned by the employed acquisitions.

In \citet{Vincendon2008} the same authors map the AOD in the SWIR
above the south seasonal cap of Mars from mid-spring to early summer.
This mapping is based on the assumption that the reflectance in the
2.64 \micron\ saturated absorption band of the \co\ ice at the
surface is mainly due to the light scattered by aerosols above most
places of the seasonal cap. In this case, one geometry is sufficient
for the AOD retrieval. Nonetheless, this method is restricted to the
area of \co\ deposits that are not significantly contaminated by
dust nor water ice. 

The Compact Reconnaissance Imaging Spectrometer for Mars (CRISM) is
a hyperspectral imager on the Mars Reconnaissance Orbiter (MRO) spacecraft.
In targeted mode, a gimbaled Optical Sensor Unit (OSU) removes most
along-track motion and scans a region of interest that is mapped at
full spatial and spectral resolution (18 or 36 m/pixel, 362-3920 nm
at 6.55 nm/channel). In the targeted mode, ten additional, spatially
binned images (180 m/pixel) are taken over the same region before
and after the main image at 10 emergence angles ranging from -70\textdegree{}\ to
70\textdegree{}. They provide the so-called Emission Phase Function
(EPF) for the site of interest that is intended for atmospheric study
and correction of atmospheric effects.

Regarding the atmospheric correction of CRISM data, \citet{McGuire2009}
adapted and improved the so-called volcano-scan technique \citep{Langevin2006}.
This method removes the \co\ gas absorption bands of any spectrum
of interest after division by a scaled reference spectrum (i.e. the
ratio between the atmospheric transmission at the summit and the base
of Olympus Mons evaluated on a Martian sol when the amounts of ice
and dust aerosols were minimal). This simple technique works reasonably
well for surfaces spectrally dominated by minerals, water ice, and
sometimes CO2 ice but does not correct for aerosol effects .

In \citet{McGuire2008}, a DISORT-based model retrieved the dust and
ice AOD, the surface pressure and temperature from previous experiment
products as well as the acquisition geometry, and the measured I/F
spectrum as inputs. Then, a surface Lambertian albedo spectrum is
computed as the output. However this algorithm does not retrieve the
AOD directly from the images nor does it take advantage of the EPF
measurements of the CRISM targeted mode.

\citet{Brown2009} proposed a first attempt in that direction by using
the DISORT algorithm to model the signal at one wavelength (i.e. 0.696
\micron). They iteratively adjust three parameters (surface albedo,
dust and ice opacity) in order to achieve a close fit at five points
spread across the EPF curve. Nevertheless the method is time consuming
and the surface albedo is assumed to be Lambertian. It has been proved
that this assumption bias the AOD and surface estimation \citep{lyapustin99a}.

In this article, we propose an original method that overcomes the
previous limitations to retrieve the optical depth of the Martian
dust from OMEGA or CRISM data at the reference wavelength of one micron.
The method is based on a parametrization of the radiative coupling
between aerosol particles and gas that determines, based on the local
altimetry and the meteorological situation, the absorption band depth
of gaseous \co. We consider the intensity of the absorption feature
at 2 \micron\ as a proxy of the AOD, provided that the other influencing
factors have been taken into account. Our method relies on radiative
transfer calculations that assume lambertian properties for the surface
even though, as demonstrated in this paper, the influence of the latter
hypothesis is minimized by considering the effect of the aerosols
on the gaseous absorption and not the total signal. When processing
OMEGA observations, we are complementary to the method of \citet{Vincendon2008}
since our approach processes pixels occupied by a wider variety of
materials - pure mineral or water ice as well as \co\ and \ho\ deposits
contaminated by a large amount of dust - while being observed at a
fixed geometry. When processing CRISM observations we take full advantage
of the top of the atmosphere spectral EPF measured by the instrument
for the retrieval of the AOD. 

This article is organized as follows. In section \ref{sec:post-proc-form}
we describe the post-processing and the formatting of the sequence
of EPF calibrated image cubes accompanying the high resolution CRISM
observation. In section \ref{sec:RT}, we give insights about Martian
atmospheric properties and radiative transfer (RT) in the SWIR range.
Furthermore we describe models that calculate the spectral radiance
coming from Mars and reaching the sensor at the top of the atmosphere.
In section \ref{sec:metho} we expose the basic assumptions, RT parametrization,
and properties on which relies the method for retrieving the AOD from
the images. In addition we describe how we implement the method that
is then validated on synthetic data. Finally, in section \ref{sec:results-select-imag},
we present and discuss results obtained for representative observations,
one from OMEGA and the others from CRISM. Conclusions are eventually
drawn in the last Section.

\section{Post-processing and formatting of CRISM EPF observations}

\label{sec:post-proc-form}

Due to the composite nature of CRISM EPF acquisitions, such data need
some post-processing prior to AOD retrieval. First of all, CRISM data
are transformed from apparent I/F units (i.e. the ratio of reflected
radiance to incident intensity of sunlight) to reflectance units.
A Lambertian surface is supposed and data are divided by the cosine
of the solar incidence angle \citep{Murchie2007}.

CRISM data are corrected for artifacts caused by non-uniformities
of the instrument, residuals of the radiometric correction or external
sources. First, hyperspectral images are corrected for striping and
spiking effects. Corrections for both distortions have been proposed
by \citet{Parente2008}. Secondly, the CRISM spectrometer is affected
by a common artifact to ``push-broom\textquotedbl{} sensors, the
so-called ``spectral smile\textquotedbl{} effect. Smile effects refers
to the artifacts originated by the non-uniformity of the instrument
spectral response along the across-track dimension, \textit{i.e.},
the horizontal axis that corresponds to the data columns.  The mitigation
of smile effects is crucial since the spectral band corresponding
to the absorption maximum of the CO$_{2}$ gas at 2$\mu$m is particularly
affected by spectral smile. We remember that the proposed AOD retrieval
method is based on this spectral feature. As a matter of fact, the
estimation of the AOD might be biased if smile effects are not addressed.
Ceamanos \textit{et al.} developed a twofold method that corrects
CRISM observations for smile by mimicking an optimal smile-free spectral
response \citep{Ceamanos2010b}. Nevertheless since uncertainties
associated to the method itself might propagate to the AOD retrievals
we conservatively restrict the use of each spatially binned image
to the central columns that own the best spectral response the so-called
``sweet spot'' columns. This decimation of the data is not a problem
because the EPF sequence provides plenty of points for each eleven
geometries.

CRISM observations are spatially rearranged to evaluate the atmosphere
optical depth of a given terrain position depending on geometry. First,
the central image that owns the highest spatial resolution is binned
by a factor of ten to match the spatial resolution of the EPF series.
After that, all images are projected onto the same geographical space.
At this point, CRISM observations showing a high overlap of the eleven
acquisitions can be discriminated. In fact, a good overlap assures
the existence of a high number of terrain units that have been sensed
from many geometries. In a single hyperspectral data set that gathers
all EPF images and the central scan (hereafter called CSP, Cube SpectroPhotometric),
each super pixel conjugated to a given terrain unit gathers up to
eleven spectra depending on the completeness of the overlap. More
details can be found in \citep{Ceamanos2013}.

\section{Mars atmospheric radiative transfer}

\label{sec:RT}

In the present section we give insights about Martian atmospheric
properties and radiative transfer (RT) in the SWIR range. Furthermore
we describe models that calculate the spectral radiance coming from
Mars and reaching the sensor at the top of the atmosphere. 

We consider the Martian atmosphere as a vertically stratified medium
with a plan parallel or spherical geometry. We divide the atmosphere
into $\approx$ 30 layers spanning the 0-100 km height range. The
lower boundary of the layer $l$ is at height $z^{l}$ above the surface.
Each layer contains a mixture of gas and solid particles and is homogeneous
in temperature $T_{l}$ {[}K{]}, pressure $P_{l}$ {[}mbar{]} and
gaseous composition $\boldsymbol{C}_{l}$ {[}ppm{]}.

\subsection{Gas}

\label{sec:RT:gas}

A dozen atmospheric absorption bands due to gaseous \co\ and marginally
to \ho\ are present in the SWIR. We calculate the transmission function
of the atmosphere along the vertical, for each pixel of our spectral
images, and as a function of wavelength.

\paragraph*{LBLRTM}

For that purpose, we employ a line-by-line radiative transfer model
(LBLRTM) \citep{Clough1995} fed by the vertical compositional and
thermal profiles predicted by the European Mars Climate Database (EMCD)
\citet{Forget2006b} for the dates, locations, and altitudes of the
observations. The LBLRTM code takes into account the vertical profile
of temperature, pressure and gaseous composition and interrogates
the spectroscopic database HITRAN (<<high-resolution transmission
molecular absorption database>>, \url{http://www.cfa.harvard.edu/hitran/})
to compute for each homogeneous layer $l$ the monochromatic absorption
coefficient $a_{l}^{\nu}$ {[}m$^{\text{2}}${]} ($\Delta\nu=10^{-4}\mathrm{cm}^{-1}$)
of the gaseous mixture for the wavenumber $\nu$. The absorption lines
typical of each molecular specie are characterized by a <<Voight>>
profile that describes at the same time the broadening by Doppler
effect and by collision. The code LBLRTM from the Atmospheric and
Environmental Research (AER) company also takes into account various
spectroscopic coupling between species (some of them not being directly
active in the optical sense like N$_{\text{2}}$ or the inert gases).
If $u_{l}$ {[}m$^{\text{-2}}${]} is the vertically integrated molecular
density in terms of number in layer $l$, the vertical monochromatic
transmission follows the Beer-Lambert law: $T^{\nu}(u_{l})=\exp\left(-a_{l}^{\nu}u_{l}\right)$

\paragraph*{Method k-correlated coefficients}

In order to take into account the fine details of the gas lines, the
final RT calculation, which will integrate the gaseous as well as
particulate absorption and scattering, should be conducted in principle
according to the maximum spectral resolution of the problem ($\Delta\nu=10^{-4}\mathrm{cm}^{-1}$).
Since we must simulate tens, even hundreds, thousand spectra with
numerically intensive codes, such a requirement cannot be satisfied
practically. Each spectral channel $k$ of the sensors that we consider
(see Introduction) presents a width $\approx$ 10 cm$^{\text{-1}}$
and thus contains hundreds of lines that cannot be considered individually.
Consequently, only a statistical approach like the one proposed by
\citet{Goody1989} offers a practical solution of the problem. Starting
from the line spectra in the spectral interval $\left[\nu_{k},\nu_{k+1}\right]$
($\nu_{k}$, being the lower bound of the current channel $k$), we
build the discrete form of the frequency distribution for the absorption:

\begin{equation}
f^{k}(a_{i})=\frac{\Delta\nu}{\nu_{k+1}-\nu_{k}}\frac{W(a_{i},a_{i}+\Delta a_{i})}{\Delta a_{i}}
\end{equation}

$a_{i}\,,\, i=1,\ldots,N$ is a partition of the interval of values
taken by the monochromatic absorption coefficient $a_{l}^{\nu}$ in
$\left[\nu_{k},\nu_{k+1}\right]$, $W(a_{i},a_{i}+\Delta a_{i})$
the number of spectral points having a value falling in the bin $i$
of lower bound $a_{i}$ and of width $\Delta a_{i}$. 

First, we choose an irregular partition that gives the same number
of points for each bin $i$. The objective is to estimate numerically
the frequency $f^{k}(a_{i})$ (i.e. a probability) with bins all containing
the same number of individuals ($\approx50)$, a number large enough
to guaranty a correct and homogeneous accuracy. Because there are
much more lines of weak intensity than lines of strong intensity,
the width $\Delta a_{i}$ is a strongly increasing function of $i$.
From $f^{k}(a_{i})$ the discrete form of the cumulative frequency
distribution $g^{k}(a)$ is derived such as: $g^{k}(a_{n})=\sum_{i=1}^{n}f^{k}(a_{i})\Delta a_{i}$.
The inverse function $a(g)=\left(g^{k}\right)^{-1}(a)$, which is
generally defined in the literature as a <<k distribution>> and
is established in the interval $g\in\left[0,1\right]$, is the highest
absorption value reached by the lower fraction $g$ of the population
of lines ordered according to their intensity. In practice, the previous
function is calculated for any value of $g$ by spline interpolation
of the experimental curve passing through the points $\left(g^{k}(a_{n}),a_{n}\right)$. 

Second, we split the line population in ($N+1$) sub-sets each representing
a fraction $\Delta g_{i}=g_{i+1}-g_{i}\,,\: i=1,\ldots,N-1$, $\Delta g_{0}=g_{1}$
and $\Delta g_{N}=1-g_{N}$ of the total. A Gaussian quadrature $g_{i}\,,\, i=1,\ldots,N$
of order N=2$^{\text{m}},\,\mathrm{m}\in\mathbb{N}$ for example allows
an optimal sampling of the weak and strong absorption values. Then,
the function $a(g)$ allows to build the corresponding partition but
in the space of absorption values $a_{i}=a(g_{i})$. We calculate
the mean value $\overline{a_{i}}$ of the monochromatic absorption
coefficient in each bin $\left[a_{i},a_{i}+\Delta a_{i}\right]$ thanks
to the values $a_{l}^{\nu}$ returned by LBLRTM AER.

Finally, as in \citet{Lacis1991}, we assume that the mean transmission
value for the instrument channel $k$ through the layer $l$ is: 

\begin{equation}
T^{k}(u_{l})=\int_{0}^{\infty}\exp\left(-au_{l}\right)f^{k}(a)da=\int_{0}^{1}\exp\left(-a(g)u_{l}\right)dg\approx\sum_{i=0}^{N+1}\exp(-\overline{a}_{i}^{l,k}u_{l})\Delta g_{i}
\end{equation}

Furthermore, we introduce a mean optical depth linked to the sub-sets
$i=0,\ldots,N+1$, each representing a fraction $\Delta g_{i}$ of
the population of lines sorted by increasing intensity : $\overline{\tau_{gaz}}_{i}^{l,k}=\overline{a}_{i}^{l,k}u_{l}$.
Then $T_{gaz}^{l,k}=\sum_{i=1}^{N+1}w_{i}\exp\left(-\overline{\tau_{gaz}}_{i}^{l,k}\right)$
with $w_{i}$ the set of coefficients associated to the Gaussian partition
$g_{i}$.

\paragraph*{Spectral transmittivity}

If we now consider the whole stack of atmospheric layers, we must
make an additional hypothesis in order to calculate an approximation
of the total vertical transmission. It is based on the fact that the
atmospheric composition does not change significantly over the heights
$z$ spanned by the region that contributes predominantly to the transmission.
Thus, the form of the distribution $a(g)$ is very similar from one
layer to the other even though the absolute level -which depends on
the pressure level for example- varies. Indeed, if an intense absorption
line is centered at wavenumber $\nu$ for layer $l$ then the line
originating from the same transition has a high probability of falling
nearly at the same position for the layer $l'\neq l$ even though
its width and intensity are different. In other words, the probability
that a strong absorption level in the layer $l$ at wave number $\nu$
is associated to a weak level in the layer $l'$ is very low. Statistically
such a situation translates to the following rule concerning the probability
that the absorption level for the wavenumber $\nu$ belongs to the
sub-set $\Delta g_{i}$ in the layer $l$ and to the sub-group $\Delta g_{j}$
in the layer $l'$ : $P\left((\Delta g_{i})^{l}\mid(\Delta g_{j})^{l'}\right)\thickapprox\delta_{ij}$
(Kronecker symbol). The transmission through both layers that should
be generally written: 
\begin{equation}
T^{k}(u_{l},u_{l'})=\sum_{i}^{N}\sum_{j}^{N}P\left((\Delta g_{i})^{l}\mid(\Delta g_{j})^{l'}\right)\, exp(-(\overline{a}_{i}^{l,k}u_{l}+\overline{a_{j}}^{l',k}u_{l'}))(\Delta g_{i})^{l}(\Delta g_{j})^{l'}
\end{equation}
 can then be approximated by : 
\begin{equation}
T^{k}(u_{l},u_{l'})\approx\sum_{i}^{N}\exp\left(-(\overline{a_{i}}^{l,k}u_{l}+\overline{a_{i}}^{l',k}u_{l'})\right)\Delta g_{i}
\end{equation}
since we use the same quadratures for both layers. This formula can
be generalized to the whole stack of layers and the vertical transmission
through the atmospheric gases in a multilayered medium is given by: 

\begin{equation}
T_{gaz}^{k}=\sum_{i}^{N+1}\Delta g_{i}\exp(-\sum_{l}\bar{\tau}_{gaz\ i}^{l,k})
\end{equation}

Because the complete calculation is very time consuming, it is only
performed for a limited ($\lesssim20$) number of reference points
representative of ``regions'' sliced according to regular bins in
latitude $lat$, longitude $long$ and altitude $h$. Three transmission
spectra are generated for the maximum, mean and minimum altitude of
a given region. Then the transmission spectrum of all the pixels belonging
to the region is interpolated from the triplet depending on their
individual altitude to give $T_{gaz}^{k}(h,lat,long)$

\subsection{Aerosols}

\label{sec:RT:aerosols}

Optical properties of Martian mineral aerosols are still largely undocumented
even though recent studies have improved our understanding \citep{Korablev2005}.
We favor the single scattering albedo, optical depth spectral shape
and phase function retrieved in the near-infrared by \citet{Vincendon2008}
and \citet{Wolff2009}. In the first case,a  Henyey-Greensten phase
function with one parameter is used. This is a coarse parametrization
but it is relevant  for the phase angle range spanned by the data
set of nadir OMEGA observations that we consider. In the second case,
because it has been proved that the Henyey-Greenstein function is
somewhat inaccurate to recreate the effect of aerosols for the CRISM
angular ranges, the radiative properties are described with more details.
A single scattering albedo, optical depth spectral shape, and phase
function based on cylindrical particles are considered. The composition
of aerosols has been shown to be remarkably homogeneous, e.g. between
Rover landing sites. Changes of radiative properties \citep{Vincendon2009a}
mainly result from differences in the mean of the aerosols size distribution
\citep{Wolff2003} because the latter results from a balance between
sedimentation and lifting mechanisms, which in turn depend on turbulence,
density, etc. Additionally the degree of hydration for the mineral
particles might change with latitude and season . Fortunately, for
the typical hydration values reached on Mars, the spectral effects
are only noticeable between 2.7 and 3.5 \micron\ \citep{Pommerol2009},
a range that we thus ignore. Water ice crystals may also be frequently
present but are not considered in this paper because they may influence
the CO$_{2}$ gas 2$\mu$m absorption feature in a specific manner.
Because the atmosphere is usually well mixed, we take a vertical distribution
of optical depth at a reference wavelength ($k_{0}$ : 1 \micron)
that is exponentially decreasing with height such that: 
\begin{eqnarray*}
\tau_{aer}^{l,k_{0}} & = & \tau_{aer}^{k_{0}}(\exp(-z^{l}/H_{scale})-exp(-z^{l+1}/H_{scale}))\\
 &  & /(1-\exp(-zmax/H_{scale}))
\end{eqnarray*}
 $\tau_{aero}^{k_{0}}=\sum_{l=1}^{L}\tau_{aer}^{l,k_{0}}$ is the
column integrated opacity of the aerosols at the reference channel
$k_{0}$, $H_{scale}$ is the scale height of the distribution, $z^{l}$
the height of layer $l$ lower interface and $zmax=100$ km. The scale
height $H_{scale}$ and the optical depth $\tau_{aero}^{k0}$ remain
the only free parameters concerning the atmosphere.

\selectlanguage{british}%

\selectlanguage{english}%

\subsection{Gas-aerosols coupling\label{sub:Couplage-gaz-aerosols}}

In general, photons undergo absorption and multiple scattering events
when interacting both with gas and aerosols. One should note that
we neglect Rayleigh scattering due to atmospheric molecules because,
according to the formula of \citet{Bodhaine1999}, that will lead
for a mean ground pressure of 6 mbar to an optical depth of $\tau_{ray}\approx5.10^{-5}$
- far lower than the contribution we can expect for the aerosols in
any case. The value of the monochromatic aerosol optical depth (at
$\Delta\nu=10^{-4}\mathrm{cm}^{-1}$) does not vary significantly
within a channel $k$ whereas the counterpart for the gas can vary
by several orders of magnitude. As a consequence calculating at channel
$k$ the transfer of radiation in the atmosphere requires to couple
gas and aerosols properties at each atmospheric level $l$ and for
each sub-set $i=0,\ldots,N+1$ of the gas line population. As indicated
in Table \ref{tab:quant_rad_atmos}, the coupling procedure leads
to $N+1$ sets of optical depth, single scattering albedo and phase
function that characterize the layers. These properties, as well as
the acquisition geometry ($\theta_{i},\theta_{e},\phi_{e}$), are
introduced one by one in a RT engine (e.g. DISORT, \citet{Stamnes1988})
in order to calculate for each level $z$ the partial radiance field
$I_{i}^{k}(z,\theta_{i},\theta_{e},\phi_{e})$ corresponding to each
sub-set lines+aerosols. The surface is characterized by its bidirectional
reflectance distribution function (BRDF). The total radiance field
is the mean value of the partial fields weighted by the fractions
$\Delta g_{i}$ : 

\begin{equation}
I^{k}(z,\theta_{i},\theta_{e},\phi_{e})=\sum_{i}^{N+1}\Delta g_{i}\, I_{i}^{k}(z,\theta_{i},\theta_{e},\phi_{e})
\end{equation}

\section{Method for retrieving the optical depth}

\label{sec:metho}

The proposed method is based on a parametrization of the radiative
coupling between aerosol particles and gas that determines, with local
altimetry and the meteorological situation, the absorption band depth
of gaseous \co. The coupling depends on (i) the acquisition geometry
(ii) the type, abundance and vertical distribution of particles (iii)
the bidirectional reflectance factor of the surface (BRF). For each
spectrum of an image, we compare the depth of the 2 \micron\ absorption
band of gaseous \co\ that we estimate on the one hand from the observed
spectrum and on the other hand from a calculated transmission spectrum
through the atmospheric gases alone. This leads to a relevant new
parameter that directly depends on $\tau_{aer}^{k0}$. Combining the
latter with the radiance factor $R_{obs}^{k_{1}}$ within the 2 \micron\ band,
we evaluate by RT inversion the AOD $\tau_{aer}^{k0}$ and the reflectance
factor of the surface.

\subsection{Parametrization of the spectral signal}

\label{sec:metho:hypothesis}

Our method is based on two main assumptions:

1. First we assume that the surface reflects the solar and atmospheric
radiation isotropically and, consequently, that the ground-atmosphere
interface is characterized by a single scalar quantity (the normal
albedo $A_{surf}^{k}$) that depends on wavelength.

2. Second we parametrize the radiative coupling between aerosols and
gas by assuming that the latter contribute to the signal as a simple
multiplicative filter. In this way, the top of the atmosphere (TOA)
radiance is written such that

\begin{equation}
I^{k}(\theta_{i},\theta_{e},\phi_{e})\approx T_{gaz}^{k}(h,lat,long)^{\epsilon^{k}(\theta_{i},\theta_{e},\phi_{e},\tau_{aer}^{k0},H_{scale},A_{surf}^{k})}I_{surf+aer}^{k}(\theta_{i},\theta_{e},\phi_{e})\label{eq:param_TR_gaz_aero}
\end{equation}

the effect of the acquisition geometry, the aerosols, and the surface
Lambertian albedo $A_{surf}^{k}$ being taken into account by scaling
the aerosol free vertical transmission $T_{gaz}^{k}(h,lat,long)$
-calculated according to \ref{sec:RT:gas}- by an exponent $\epsilon$.
\\
 $I_{surf+aer}^{k}(\theta_{i},\theta_{e},\phi_{e})$ is the calculated
TOA radiance but without considering the atmospheric gases in the
radiative transfer. Factor $\epsilon^{k}$ can be further decomposed
into two terms: 

\begin{equation}
\epsilon^{k}(\theta_{i},\theta_{e},\phi_{e},\tau_{aer}^{k0},H_{scale},A_{surf}^{k})=\psi(\nu)\beta^{k}(\theta_{i},\theta_{e},\phi_{e},\tau_{aer}^{k0},H_{scale},A_{surf}^{k})\label{eq:coef_epsilon}
\end{equation}

with the airmass $\nu=\frac{1}{\cos(\theta_{i})}+\frac{1}{\cos(\theta_{e})}$.
Factor $\psi$ is purely geometric and allows a quick and simplified
calculation of the free gaseous transmission along the acquisition
pathlength (incident-emergent) using the vertical free transmission
: 

\begin{equation}
T{}_{gaz}^{'k}=\sum_{i}^{16}a_{i}\exp(-\sum_{l}\bar{\tau}_{gaz\ i}^{l,k}.\nu)\approx(T_{gaz}^{k})^{\psi(\nu)}\label{eq:trans_gas_param}
\end{equation}

the approximate formula being $\psi(\nu)=p_{0}(p_{1}+p_{2}\nu+p_{3}\nu^{2}+p_{4}\nu^{3})$.
In order to retrieve the values of the factors $p_{1}$ to $p_{4}$,
we fit the experimental cloud of points obtained by plotting $\nicefrac{log\left(T{}_{gaz}^{'k}\right)}{log(T_{gaz}^{k})}$
as a function of the airmass $\nu$ for a set of representative atmospheric
vertical profiles. For that purpose the quantities $T{}_{gaz}^{'k}$
and $T_{gaz}^{k}$ are respectively the slant and vertical versions
of the free transmission through the gas given by the LBLRTM code.\textbf{
}Note that the dispersion around the mean curve is very limited. .
Factor $p_{0}$ accounts for the difference of spectral resolution
and radiometric calibration between OMEGA and CRISM and its retrieval
is explained at the end of section \ref{sub:beta-inversion}. All
p-factor values are given in Table \ref{tab:coef_psi}.

Factor $\beta^{k}$ on the other hand expresses the aerosol effect
on gaseous absorption and depends on the considered channel $k$.
In practice this spectral dependability can be easily managed. Indeed,
numerical experiments and the analysis of real data (see following
sections) show that this exponent depends principally on the gaseous
absorption intensity $T_{gaz}^{k}$. If (i) $\beta^{k}$ can be estimated
experimentally ($\hat{\beta^{k}}$) for one or for several geometries
for each pixel (super-pixel) of an OMEGA (CRISM) observation and (ii)
if one assumes a value for $A_{surf}^{k1}$ ($k_{1}$ is the spectral
band number at 2\micron) then $(\tau_{aer}^{k0})$ can be derived.
Indeed, function $\beta^{k}(\theta_{i},\theta_{e},\phi_{e},\tau_{aer}^{k0},H_{scale},A_{surf}^{k_{1}})$
is invertible provided that the scale height $H_{scale}$ of the aerosols
is known. In this matter, several studies such as \citet{Vincendon2008}
suggest that $H_{scale}=11$ km is a good guess, but in section \ref{sec:results-select-imag}
we prove that $H_{scale}=8$ km may be more appropriate as it leads
to better results.

\subsection{Numerical aspects LUT\label{sec:metho:numerical-aspects}}

\begin{figure}
\includegraphics[width=0.9\textwidth]{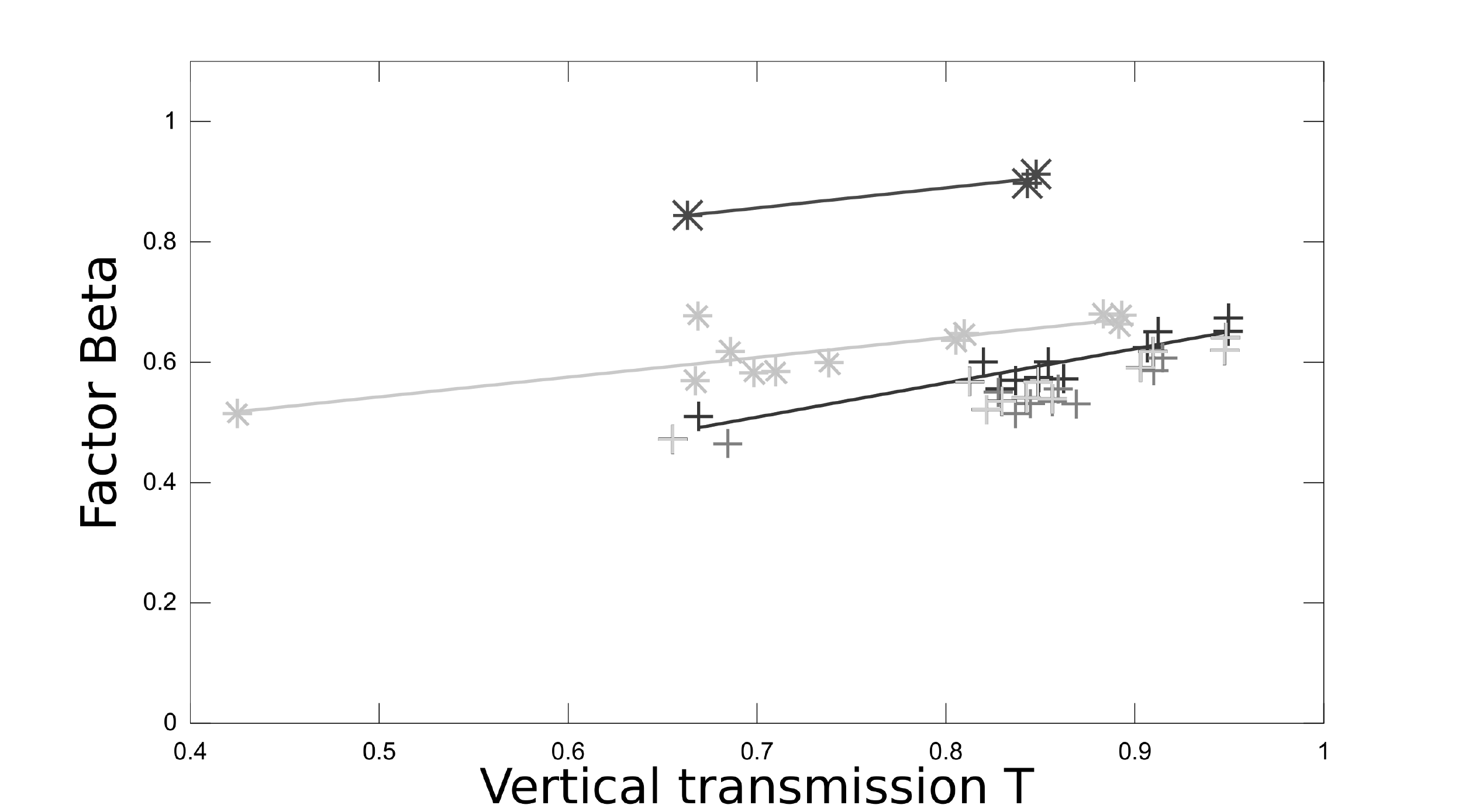}

\caption{Experimental scatter-plots between $T_{ref}^{k}$ and $\beta^{k}$
for different geometries and atmospheric conditions. The  points represented
by the {*} symbols and fitted by regression lines of the same grey
level, come from synthetic data. The points represented by the + symbols
and fitted by a single regression line  come from the estimation of
factor $\beta^{k}$ for three different observed spectra, each of
them corresponding to a specific grey level (see section \ref{sec:metho:estim_beta})
\label{fig:courbe_T-beta}}
\end{figure}

Our optical depth retrieval system uses a multi-dimensional look-up
table (LUT) of $\beta$ values arranged according to discrete combinations
of acquisition geometries $(\theta_{i},\theta_{e},\phi_{e})$ and
of physical parameters $(\tau_{aer}^{k0},H_{scale},A_{surf}^{k1})$,
see Table \ref{tab:LUT-beta}. For each combination, the value of
$\beta$ is calculated on the basis of three TOA spectra. $T_{ref}^{k}$
is a reference free vertical transmission corresponding to a given
time and location on the surface of Mars. It has been chosen in order
to cover a range of absorption values (all spectral bands considered)
typical of the southern hemisphere in spring. The spectra $I^{k}$
and $I_{surf+aer}^{k}$ in units of radiance are generated successively
by the code DISORT with and without considering the gaseous absorption
in the atmosphere. A series of k-correlated coefficients $\bar{\tau}_{ref\ i}^{l,k}$
intervenes in the calculation of $T_{ref}^{k}$ and $I^{k}$. Finally
the LUT is built according to the formula: 

\begin{equation}
\beta_{ref}(\theta_{i},\theta_{e},\phi_{e},\tau_{aer}^{k0},H_{scale},A_{surf}^{k1})=\sum_{k=k_{1}'}^{k_{1}''}f_{k}\beta^{k}=\sum_{k=k_{1}'}^{k_{1}''}f_{k}\ln(\frac{I^{k}}{I_{surf+aer}^{k}})/\ln(T{}_{ref}^{'k})
\end{equation}
 with $f_{k}=1/K$. The channels $k=k'_{1},\ldots,k''_{1}$ ($K=k''_{1}-k'_{1}+1$)
encompass the 2 \micron\ \co\ gas absorption band such as $T_{ref}^{k}<0.85$.
For DISORT calculations, we consider the atmosphere as a plan parallel
vertically stratified medium which is a sufficient approximation to
calculate the TOA radiance whenever incidence and emergence directions
are $\approx$10$^{o}$ above the horizon. Several experimental scatter-plots
$T_{ref}^{k}$ - $\beta^{k}$ built using the reference and also other
sets of k-correlated coefficients show that there exists a linear
relationship between the two quantities with an excellent correlation
factor ($\gtrsim0.9$ see figure \ref{fig:courbe_T-beta}) :

\begin{equation}
\beta^{k}=a(\nu,\tau_{aer}^{k0},H_{scale},A_{surf}^{k1})+b(\nu,\tau_{aer}^{k0},H_{scale},A_{surf}^{k1})T_{ref}^{k}
\end{equation}

We notice that the coefficients $(a,b)$ of the regression depend
weakly on the parameters $\tau_{aer}^{k0},H_{scale},A_{surf}^{k1}$. 

\footnotetext[1]{the albedo of the surfaces that we consider rarely exceeds 0.6}

\subsection{Estimation of the coupling factor gas-aerosols \label{sec:metho:estim_beta}}

Because the intensity of the CO$_{2}$ gas 2$\mu$m absorption feature
is diversely accessible from the spectra depending on the nature of
the surface materials present in the pixel, we need to consider different
strategies for estimating the factor $\beta$.

\paragraph*{Mineral surfaces: method 1}

The factor $\beta$ can be readily estimated for every OMEGA or CRISM
spectrum that does not show \foreignlanguage{american}{any features,
such as the} \ho\ or \co\ ice signature, superimposed on the 2
\micron\ \co\ gas absorption band. In this way, this absorption
band is completely due to the atmosphere. A similar formula to the
one used in the volcano scan technique \citep{McGuire2009} is used
replacing the Olympus reference transmission spectrum by $(T_{gaz}^{k})^{\psi(\nu)}$
such that

\begin{equation}
\begin{array}{c}
\hat{\beta^{k}}=\frac{\alpha}{\psi(\nu)\ln\left(\frac{T_{gaz}^{k}}{T_{gaz}^{k_{3}}}\right)}\\
\alpha=\ln\left(\frac{R_{obs}^{k}}{R_{obs}^{k_{3}}}\right)+0.0909\:\ln\left(\frac{R_{obs}^{k_{0}}}{R_{obs}^{k_{3}}}\right)
\end{array}
\end{equation}

with $k_{3}$ the index of a channel falling at the shortwave extremity
of the 2 \micron\ band wing. The scatter-plot $T_{gaz}^{k}$-$\hat{\beta}^{k}$,
$k=k'_{1},\ldots,k''_{1}$ built for each pixel of a real OMEGA observation
confirms the existence of a linear relationship between the two quantities
with a generally moderate dispersion (see Figure \ref{fig:courbe_T-beta}).
The regression coefficients $(\hat{a},\hat{b})$ are then calculated
and will allow us to estimate exponent $\beta$ in the same absorption
range than the reference :

\begin{equation}
\hat{\beta}_{ref}=\hat{a}+\hat{b}\sum_{k=k_{1}'}^{k_{1}''}f_{k}T_{ref}^{k}
\end{equation}

Note that we use the series of channels $k=k'_{1},\ldots,k''_{1}$
for the estimation of $\hat{\beta}_{ref}$ allowing us to be less
sensitive to noise and channel aging than using the single channel
$k_{1}$. In order to perform the inversion process, we take as an
initial value for the lambertian albedo of the surface $A_{surf}^{k_{1}}=\frac{R_{obs}^{k_{1}}}{(T_{gaz}^{k_{1}})^{\psi(\nu)\hat{\beta}^{k_{1}}}}$
.

\paragraph{Water ice surfaces: method 2}

A specific procedure is necessary for any spectrum marked by the signature
of water ice with the possible presence of dust but without spectral
contamination by \co\ ice. A numerical optimization is performed
regarding a cost function that depends on $\beta$ and that expresses
the quality of the gaseous absorption correction on the spectra. The
quality criteria we choose is the band shape of solid \ho\ at 2
\micron\ that must show a unique local minimum and a simple curvature.
Mathematically this criteria is equivalent to the second derivative
of the spectrum that must be close to zero for channels $\left[k'_{1},k''_{1}\right]$
covering the \ho\ band and slightly beyond. After all the cost function
is written:

\begin{equation}
\xi(\beta)=\sum_{k=k'_{1}}^{k''_{1}}\left(\frac{d^{2}}{dk^{2}}\frac{R_{obs}^{k}}{(T_{gaz}^{k})^{\psi(\nu)\beta}}\right)^{2}
\end{equation}

Searching for the global minimum of this function leads to the estimation
of the exponent $\beta$ : $\hat{\beta}=\underset{\beta}{\arg\min\:}\xi(\beta)$.
The normalization of $\hat{\beta}$ up to the reference level of absorption
uses the regression line:

\[
\hat{\beta_{ref}}=\hat{b}(\sum_{k=k_{1}'}^{k_{1}''}f_{k}T_{ref}^{k}-T_{gaz}^{k_{1}})+\hat{\beta}
\]

Once again, in order to realize the inversion we take as initial value
for the lambertian albedo $A_{surf}^{k_{1}}=\frac{R_{obs}^{k_{1}}}{(T_{gaz}^{k_{1}})^{\psi(\nu)\hat{\beta}}}$.

\paragraph{Carbon dioxide ice (\co)}

Currently there is no simple strategy to estimate in a pixelwise manner
the $\beta$ coefficient from spectra presenting the signature of
solid \co\ because the latter cannot be distinguished from its gaseous
counterpart at 2\micron. For this kind of surface, our method to
estimate $\tau_{aer}^{k0}$ is ineffective. 

If, on the one hand, \co\ ice proves to be sufficiently pure with
grain sizes more than 100 \micron\ in diameter (this conditions is
always respected in practice for the seasonal deposits) then the method
proposed by \citep{Vincendon2008} is the only solution. It is based
on the assumption that the reflectance level at 2.63 \micron\ (channel
$k_{2}$) is nearly equal to zero because of the absorption saturation
of the ice. Consequently, in this case, the reflectance factor $R_{obs}^{k_{2}}$
represents the path radiance of the atmosphere (mostly due to the
aerosols at channel $k_{2}$) and is a bijective, easily invertible
function of $\tau_{aer}^{k0}$. In section \ref{sub:beta-inversion}
we will see how exponent $\beta$ can be calculated from $\tau_{aer}^{k0}$,
computed using \citep{Vincendon2008}, and $R_{obs}^{k_{1}}$ for
correcting the gaseous absorption on the spectra (section \ref{sub:beta-inversion}). 

If, on the other hand, \co\ ice proves to be substantially contaminated
by dust or water ice, then the only conceivable way to estimate $\beta$
could be a statistical procedure as it is proposed in \citet{BinLuo2010}.

Sulfates\textbf{ }present a broad band similar to water ice around
2 \micron\  and are occasionally spectrally dominant in OMEGA or
CRISM observations. A future requirement is finding a way to deconvolve
the CO\textsubscript{2} gaseous features from the sulfates 1.94 strong
absorption in order to evaluate $\beta$ .

Hereafter, we simplify the writing of the parameters by removing subscript
$ref$ from the notations. The quantities $\beta$ and $\hat{\beta}$
are as if they were wavelength independent because they are normalized
up to the reference level of absorption.

\subsection{Sensitivity and feasibility study. \label{sub:Sensitivity-and-feasibility}}

\paragraph*{Synthetic data}

We now examine the sensitivity of function $\beta$ regarding its
different parameters $\theta_{i},\theta_{e},\phi_{e},\tau_{aer}^{k0},H_{scale},A_{surf}^{k1}$.
This study is summarized as follows. The exponent $\beta$ decreases
with the airmass $\nu$ especially as the aerosol opacity is on the
rise. This is the direct consequence of an intensifying coupling effect
between gas and aerosols, the latter reducing the effective pathlength
of the photons in the atmosphere by scattering and thus the probability
to be absorbed by \co\ gas. The exponent evolution reflects such
a decrease of relative absorption band intensity. The coupling is
higher when the sensor looks in the solar direction $(90\lesssim\phi_{e}\lesssim180\text{\textdegree) }$
than when its looks in the opposite half space $(0\lesssim\phi_{e}\lesssim90\text{\textdegree) }$.
The higher the scale height characterizing the vertical distribution
of the aerosols, the more aerosols influence atmospheric absorption
especially since their opacity is high. The surface reflectance factor
also controls the absorption by \co\ gas because of occurrence of
multiple scattering between the two media. A bright surface promotes
a higher number of roundtrips for the photons which then have a higher
probability to be absorbed: exponent $\beta$ increases with Lambertian
albedo $A_{surf}^{k1}$. Finally, because the sensitivity of $\beta$
with $\tau_{aer}^{k0}$ ($\frac{\partial\beta}{\partial\tau}$) decreases
with the airmass, we can expect significant estimation errors for
$\tau_{aer}^{k0}$ below a certain threshold. Indeed the ground pressure
provided by the Martian climate database \citep{Forget2006b} is only
indicative of the real pressure at the moment of the observation.
We must stress that the ground pressure determines $T_{gaz}^{k}$
then $\hat{\beta}$ and, for this reason, an uncertainty $\Delta p_{surf}$
will propagate as an uncertainty $\Delta\beta$ on the coupling factor,
then as an uncertainty on the optical depth $\Delta\tau_{aer}^{k0}=\frac{\partial\tau}{\partial\beta}\Delta\beta$
. We quantify this effect in the validation section \ref{sub:Validation}.

The experimental behavior of $\hat{\beta}$, evaluated above mineral
surfaces based on real data with respect to airmass $\nu$ and $\tau_{aer}^{k0}$,
is also full of information. We distinguish two situations according
to whether we work with OMEGA or CRISM observations.

\paragraph*{OMEGA data}

\begin{figure}
\includegraphics[bb=0bp 150bp 550bp 650bp,clip,width=0.7\paperwidth]{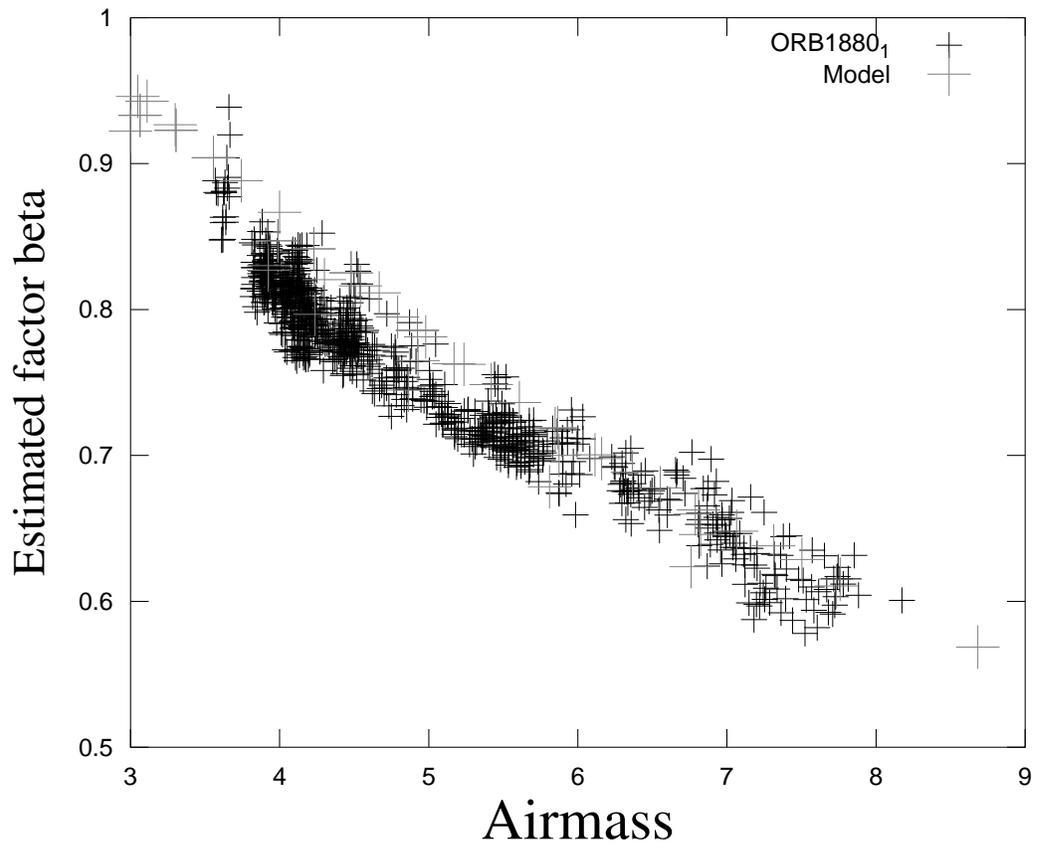}

\caption{Scatter plot \textbf{(}$\mathrm{airmass}\:\nu,\hat{\beta}$) for the
observation 1880\_1 related to the modeled ($\nu,\beta$) for $H_{scale}\approx11$
km, $\tau_{aer}^{k0}\approx0.8$ and $A_{surf}^{k1}\approx0.2$. \label{fig:plot_airmass_beta_OMEGA}}
\end{figure}

In the OMEGA case, we consider large populations of pixels extracted
from several hyperspectral images. The principal variability of the
aerosol opacity across the scene comes from altitude changes -sometimes
several kilometers in elevation- that modify the local atmospheric
height and thus the column integrated density of the aerosols. The
horizontal heterogeneity of the particle density at constant altitude
generally implies a more moderate variability. Hence, by selecting
pixels belonging to slices spanning no more than a couple hundred
meters in elevation but observed under various geometries, we can
draw a scatter-plots (\textbf{$\nu,\hat{\beta}$}) such as that presented
in Fig. \ref{fig:plot_airmass_beta_OMEGA}. Set apart a noticeable
internal dispersion linked to factors such as those previously mentioned
or spatial variations of Lambertian surface albedo, the points corresponding
to the OMEGA observation 1880\_1 are organized according to monotonously
decreasing curves of $\hat{\beta}$ over a wide range of increasing
$\nu$ values. The curves faithfully follow the theoretical scatter
plot that can be built thanks to the numerical reference LUT for $H_{scale}\approx$
11 km and $A_{surf}^{k1}\approx0.2$. The best fit gives  an estimation
of the mean optical depth $\tau_{aer}^{k0}\approx$ 0.8. 

\paragraph*{CRISM data}

The sensitivity study is conducted on CRISM EPF sequence FRT82EB.
This observation corresponds to an ice-free cratered area of the south
pole of Mars ($lat$=-83$^{\circ}$). The image geometry corresponds
to the conditions of the high latitudes with $\theta_{i}$=74.5$^{\circ}$.
Due to the sun-synchronous orbit of CRISM, there are two modes in
azimuthal angle, $\phi_{e}\simeq$34$^{\circ}$ and $\phi_{e}\simeq$146$^{\circ}$.
Regarding the topography, FRT82EB presents mild slopes and a constant
altitude. The initial albedo of the surface can be approximated by
the average value of the spectral band at 1 micron. In this case,
the $A_{surf}^{k1}$ of FRT82EB is equal to 0.3.

Data are processed by the data pipeline described in Section \ref{sec:post-proc-form},
except the filtering of the columns affected by the ``spectral smile''
effect.  Thanks to the good overlap of the high-resolution image and
of the EPF, we can define up to 850 super pixels that are represented
by more than 6 geometries.

\begin{figure}[t]
\includegraphics[width=0.5\paperwidth]{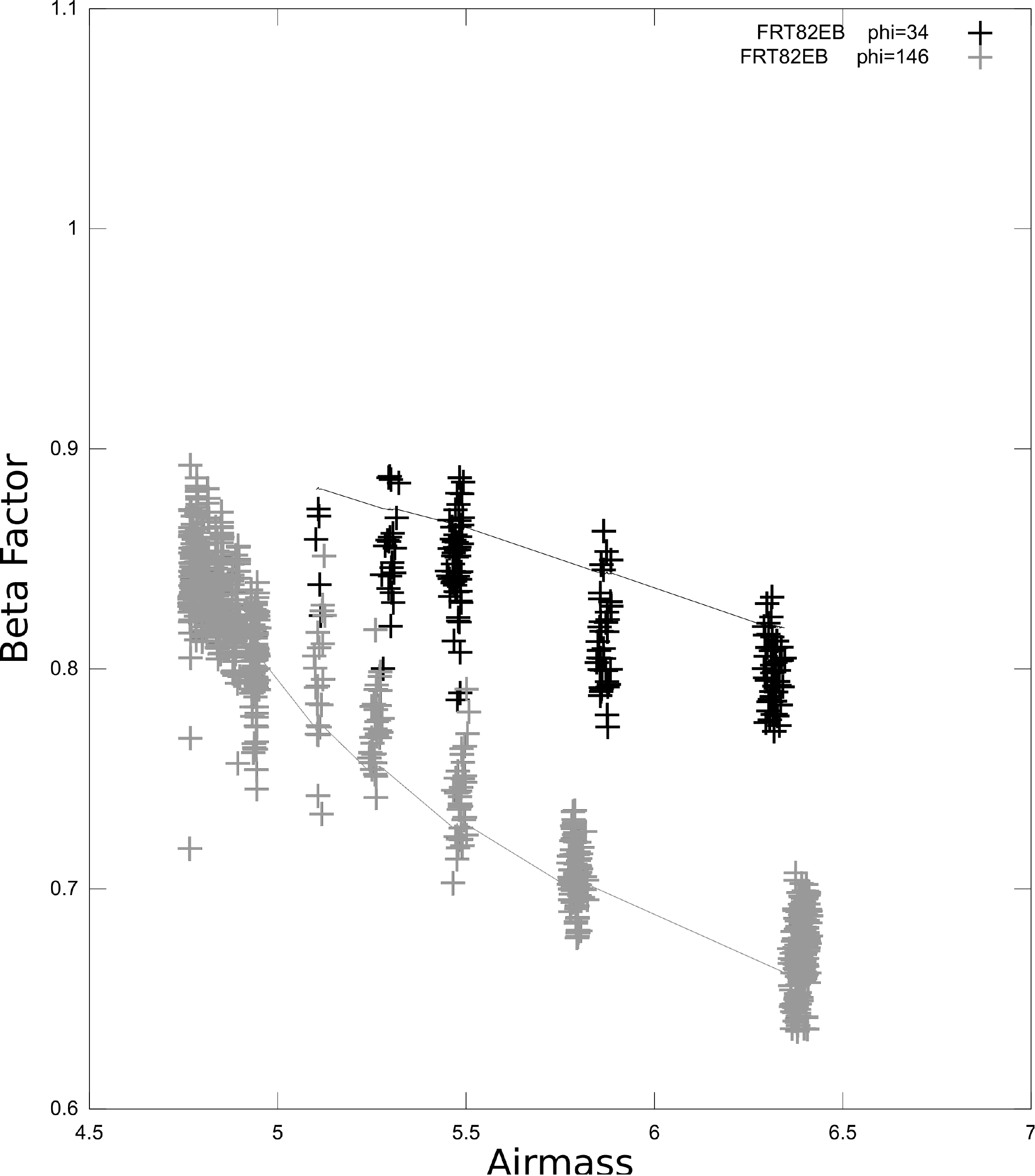}
\caption{Experimental $\beta$ curves from the CRISM image FRT82EB according
to $airmass$. The model curves that provide the best match are also
plotted as solid lines. Note the dispersion of $\beta$ values likely
linked with the ``spectral smile'' affecting the spatially binned
images, i.e. the clusters of points in the scatter plot. }

\label{fig:beta} 
\end{figure}

The factor $\beta$ can be readily estimated for every CRISM spectrum
that does not show \co\ ice signature superimposed on the 2 \micron\ \co\ gas
absorption band. Once a series of spectra corresponding to a super
pixel is treated, we obtain $\beta$ as a function of up to eleven
geometries. Figure \ref{fig:beta} summarizes the results for the
image FRT82EB by plotting $\beta$ as the function of the airmass
distinctively for the two azimuths explored by the observation (black
and grey crosses). 

\begin{figure}
\includegraphics[width=0.5\paperwidth]{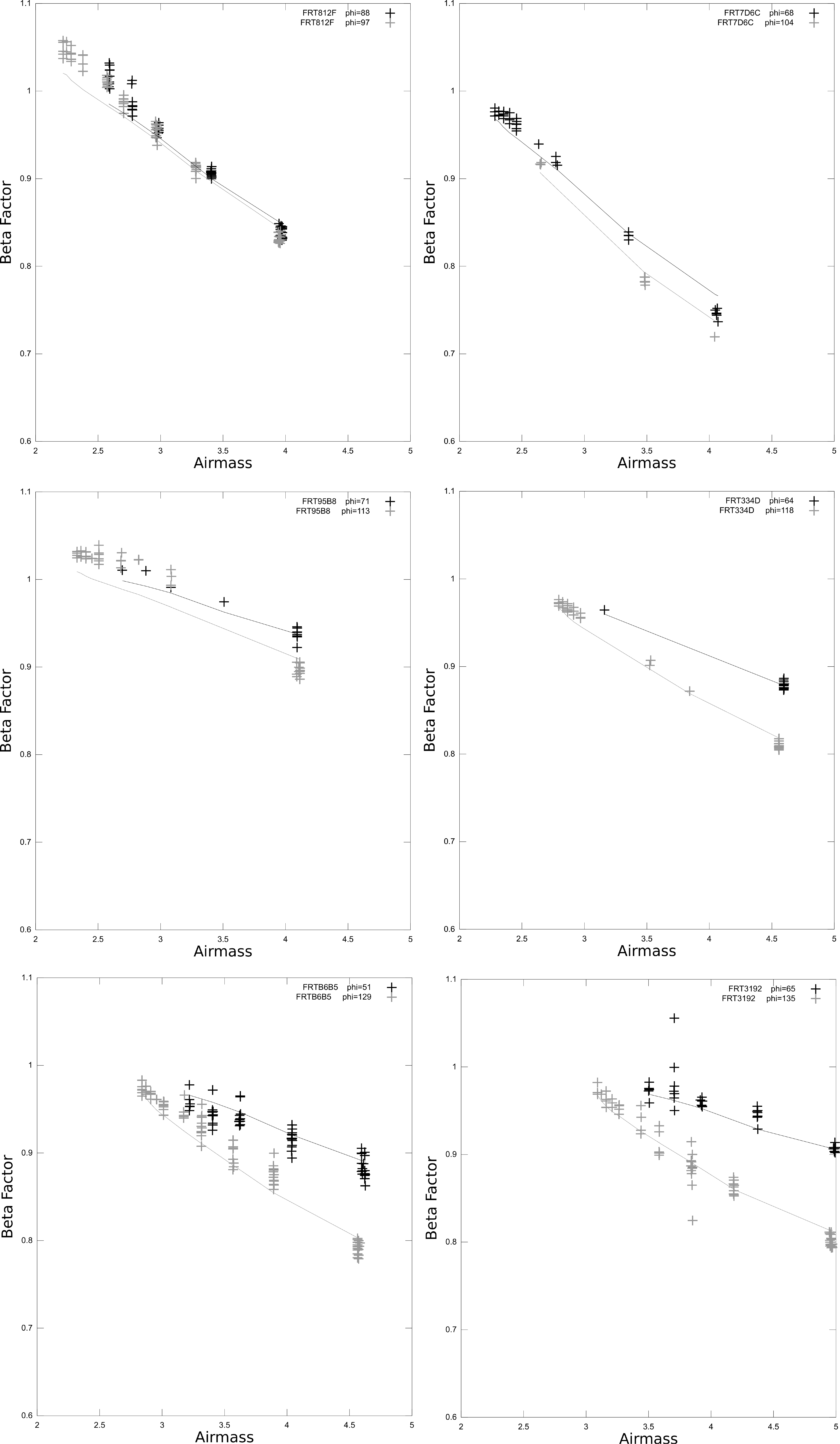}

\caption{Experimental $\beta$ curves from CRISM selected EPF sequences as
a function of airmass plotted distinctively for the two azimuths explored
by the observation (black and grey crosses). In each case, the model
curves that provide the best match are also plotted as solid lines.\label{fig:beta-curves-FRT-Gusev}}
\end{figure}

One can see how the azimuthal angle has an impact on the gas-aerosol
coupling due to aerosol phase function strong anisotropy \citep{Wolff2009}.
In addition, the model curves of $\beta$ that provide the best matching
with the experimental results are shown by plain lines. In fact, the
best match corresponds to a value of $\tau_{aer}^{k0}$ around  0.5,
respectively 8 km for $H_{scale}$. Curves for other targeted observations
are shown in Fig. \ref{fig:beta-curves-FRT-Gusev}.

\subsection{Algorithms for the simultaneous retrieval of surface albedo and aerosol
opacity: \label{sub:beta-inversion}}

We now describe the practical implementation of the method that allows
the computation of surface albedo and aerosol opacity maps from a
given observation. The implementation comes in two flavors depending
on the origin of the data.

\paragraph*{OMEGA}

The processing of a nadir OMEGA image is conducted on a pixelwise
basis after segmenting the image according to an automatic detection
of the surface components \citet{Schmidt2007}. For the pixels that
do not present the signatures of solid \co\ the following iterative
algorithm is carried out:
\selectlanguage{french}%
\begin{quotation}
\begin{minipage}[t]{1\columnwidth}%
\selectlanguage{english}%
\textbf{Input}: a spectrum $R_{obs}^{k}$ of rank $p$ in the flattened
image, the corresponding vertical transmission $T_{gaz}^{k}$ calculated
as explained in section \ref{sec:RT:gas}, the acquisition geometry
$(\theta_{i},\theta_{e},\phi_{e})$ and a global estimation of the
scale height $H_{scale}$
\begin{enumerate}
\item estimating the exponent $\beta$

\begin{enumerate}
\item using method 1 (section \ref{sec:metho:estim_beta}) for a mineral
surface
\item using method 2 for a water ice surface without \co\ ice contamination
\end{enumerate}
\item initialization $i=0$, Lambertian albedo$\left(A_{surf}^{k_{1}}\right)_{0}=\frac{R_{obs}^{k_{1}}}{(T_{gaz}^{k_{1}})^{\psi(\nu)\hat{\beta}}}$
\item DO
\item \hspace*{0.35cm}building a 1D curve $\tau_{aer}^{k0}$ as a function
of $\beta$ by quadrilinear interpolation of the LUT described in
section \ref{sec:metho:estim_beta} for a given set $(\theta_{i},\theta_{e},\phi_{e})$,
$A_{surf}^{k_{1}}=\left(A_{surf}^{k_{1}}\right)_{i}$ and $H_{scale}$.
\item \hspace*{0.35cm}linear interpolation of the previous curve at $\hat{\beta}$
to calculate $\left(\tau_{aer}^{k0}\right)_{i+1}$
\item \hspace*{0.35cm}building a 1D curve $A_{surf}^{k_{1}}$ as a function
of $R_{TOA}^{k_{1}}$ by quadrilinear interpolation of the LUT for
a given set of $(\theta_{i},\theta_{e},\phi_{e})$ and $\tau_{aer}^{k0}=\left(\tau_{aer}^{k0}\right)_{i+1}$
\item \hspace*{0.35cm}linear interpolation of the previous curve at $\frac{R_{obs}^{k_{1}}}{(T_{gaz}^{k_{1}})^{\psi(\nu)\hat{\beta}}}$
in order to calculate $\left(A_{surf}^{k_{1}}\right)_{i+1}$
\item \hspace*{0.35cm}$i=i+1$
\item UNTIL $\left(\left|\left(A_{surf}^{k_{1}}\right)_{i}-\left(A_{surf}^{k_{1}}\right)_{i-1}\right|\leq0.01\:\mathrm{AND}\left|\left(\tau_{aer}^{k0}\right)_{i}-\left(\tau_{aer}^{k0}\right)_{i-1}\right|\leq0.01\right)\:\mathrm{OR}\: i\geq imax$ \selectlanguage{french}%
\end{enumerate}
\end{minipage}
\end{quotation}
\selectlanguage{english}%
The algorithm usually converges after a dozen iterations and leads
to a complete solution once $A_{surf}^{k_{0}}$ is calculated from
$R_{obs}^{k_{0}}$ and $\tau_{aer}^{k0}$ based on a dedicated LUT.
If the initial estimation $\hat{\beta}$ or, to a lesser extent, the
estimation of$\left(A_{surf}^{k_{1}}\right)_{0}$ is poor then the
convergence is not reached and there is no solution. 

For the pixels presenting a signature of solid \co\ pure enough (according
to the criteria established by \citet{Vincendon2008}) we have seen
in section \ref{sec:metho:estim_beta} that it is possible to obtain
$\tau_{aer}^{k0}$ by their independent method. The problem is then
laid down differently: we must calculate the exponent $\beta$ from
$\tau_{aer}^{k0}$ and $R_{obs}^{k_{1}}$ that will be used for correcting
the gaseous absorption on the spectra. The algorithm is also iterative
in this case :
\selectlanguage{french}%
\begin{quotation}
\begin{minipage}[t]{1\columnwidth}%
\selectlanguage{english}%
\textbf{Input} : a spectrum $R_{obs}^{k}$ of rank $p$ in the flattened
image, the corresponding vertical transmission $T_{gaz}^{k}$ extracted
from the ATM cube at the same rank, the acquisition geometry $(\theta_{i},\theta_{e},\phi_{e})$
and a global estimation of the scale height $H_{scale}$
\begin{enumerate}
\item estimation of $\tau_{aer}^{k0}$ using the method by \citet{Vincendon2008}
\item initialization $i=0$, Lambertian albedo$\left(A_{surf}^{k_{1}}\right)_{0}=0.01$
\item DO
\item \hspace*{0.35cm}building a 1D curve $\beta$ as a function of $\tau_{aer}^{k0}$
by quadrilinear interpolation of the LUT described in section \ref{sec:metho:estim_beta}
for a given set $(\theta_{i},\theta_{e},\phi_{e})$, $A_{surf}^{k_{1}}=\left(A_{surf}^{k_{1}}\right)_{i}$
and $H_{scale}$.
\item \hspace*{0.35cm}linear interpolation of the previous curve at $\tau_{aer}^{k0}$
to calculate $\left(\hat{\beta}\right)_{i+1}$
\item \hspace*{0.35cm}calculation of $C_{obs}^{1}=\frac{R_{obs}^{k_{1}}}{(T_{gaz}^{k_{1}})^{\psi(\nu)\left(\hat{\beta}\right)_{i+1}}}$ 
\item \hspace*{0.35cm}building a 1D curve $A_{surf}^{k_{1}}$ as a function
$R_{TOA}^{k_{1}}$ by quadrilinear interpolation of the LUT for a
given set $(\theta_{i},\theta_{e},\phi)$ and $\tau_{aer}^{k0}$
\item linear interpolation of the previous curve at $C_{obs}^{1}$ in order
to calculate $\left(A_{surf}^{k_{1}}\right)_{i+1}$
\item \hspace*{0.35cm}$i=i+1$
\item UNTIL $\left(\left|\left(A_{surf}^{k_{1}}\right)_{i}-\left(A_{surf}^{k_{1}}\right)_{i-1}\right|\leq0.01\:\mathrm{AND}\left|\left(\hat{\beta}\right)_{i}-\left(\hat{\beta}\right)_{i-1}\right|\leq0.01\right)\:\mathrm{OR}\: i\geq imax$ \selectlanguage{french}%
\end{enumerate}
\end{minipage}
\end{quotation}
\selectlanguage{english}%

\paragraph*{CRISM}

The processing is applied to the single hyperspectral data set that
gathers the central scan and the EPF images, i.e. the CSP cube. We
choose an area of the image that offers the best compromise between
its size and its spectral homogeneity. In addition, this area should
not present the signatures of solid \co. All the super-pixels that
fall in the area are blended into one vector of parameters that needs
to be explained as a whole by the model provided the solution for
parameters $\tau_{aer}^{k0}$ and $R_{obs}^{k_{1}}$ is found. The
following iterative algorithm is carried out.
\selectlanguage{french}%
\begin{quotation}
\begin{minipage}[t]{1\columnwidth}%
\selectlanguage{english}%
\textbf{Input} : the collection of valid spectra $R_{obs}^{k}$, the
corresponding vertical transmissions $T_{gaz}^{k}$ extracted from
the ATM cube at the same coordinates, the acquisition geometries $(\theta_{i},\theta_{e},\phi_{e})$
and a global estimation of the scale height $H_{scale}$
\begin{enumerate}
\item for each valid pixel $p$, estimating the exponent $\beta$

\begin{enumerate}
\item using method 1 (section \ref{sec:metho:estim_beta}) for a mineral
surface
\item using method 2 for a water ice surface without \co\ ice contamination
\end{enumerate}
\item initialization $i=0$, mean Lambertian albedo for the selected area$\left(\overline{A_{surf}^{k_{1}}}\right)_{0}$
\item DO
\item \hspace*{0.35cm}building a matrix of $\beta$ values, each element
$\beta^{pq}$ corresponding to the coupling factor that is predicted
by the model for the spectrum of index $p$ assuming that the atmospheric
opacity is$\left(\tau_{aer}^{k0}\right)_{q},\: q=1,\ldots,Q$, a sampling
of the range of possible variation for this parameter. The value $\beta^{pq}$
is calculated by quadrilinear interpolation of the LUT described in
section \ref{sec:metho:estim_beta} for a given set $(\theta_{i},\theta_{e},\phi_{e})$,
$A_{surf}^{k_{1}}=\left(\overline{A_{surf}^{k_{1}}}\right)_{i}$,
$\left(\tau_{aer}^{k0}\right)_{q}$, and $H_{scale}$.
\item \hspace*{0.35cm}finding the global minimum of the objective function
$\chi(\tau_{aer}^{k0})=\sum_{p=1,\ldots,P}\left(\hat{\beta}^{p}-\beta^{pq}\right)^{2}$to
estimate $\left(\tau_{aer}^{k0}\right)_{i+1}$
\item \hspace*{0.35cm}for each valid pixel $p$, building a 1D curve $A_{surf}^{k_{1}}$
as a function of $R_{TOA}^{k_{1}}$ by quadrilinear interpolation
of the LUT for a given set of $(\theta_{i},\theta_{e},\phi_{e})$
and $\tau_{aer}^{k0}=\left(\tau_{aer}^{k0}\right)_{i+1}$. Linear
interpolation of the previous curve at $\frac{R_{obs}^{k_{1}}}{(T_{gaz}^{k_{1}})^{\psi(\nu)\hat{\beta}}}$
in order to calculate $\left(A_{surf}^{k_{1}}\right)_{i+1}$
\item \hspace*{0.35cm}calculation of the mean Lambertian albedo $\left(\overline{A_{surf}^{k_{1}}}\right)_{i+1}$
for the selected area
\item \hspace*{0.35cm}$i=i+1$
\item UNTIL $\left(\left|\left(\overline{A_{surf}^{k_{1}}}\right)_{i}-\left(\overline{A_{surf}^{k_{1}}}\right)_{i-1}\right|\leq0.01\:\mathrm{AND}\left|\left(\tau_{aer}^{k0}\right)_{i}-\left(\tau_{aer}^{k0}\right)_{i-1}\right|\leq0.01\right)\:\mathrm{OR}\: i\geq imax$ \selectlanguage{french}%
\end{enumerate}
\end{minipage}
\end{quotation}
Note \foreignlanguage{english}{that the value of factor $p_{0}$ appearing
in Table \ref{tab:coef_psi} is adjusted for the OMEGA (respectively
CRISM) dataset by maximizing the overall convergence rate achieved
by algorithms 3.5a (respectively 3.5c) for a selection of observations.
Note that this optimization problem is satisfactorily convex and thus
its solution satisfactorily constrained, since the convergence of
the iterative inversion is very sensitive to the value of $p_{0}$. }

\selectlanguage{english}%

\subsection{Validation\label{sub:Validation}}

\begin{figure}
\includegraphics[bb=0bp 150bp 550bp 642bp,clip,width=1.\textwidth]{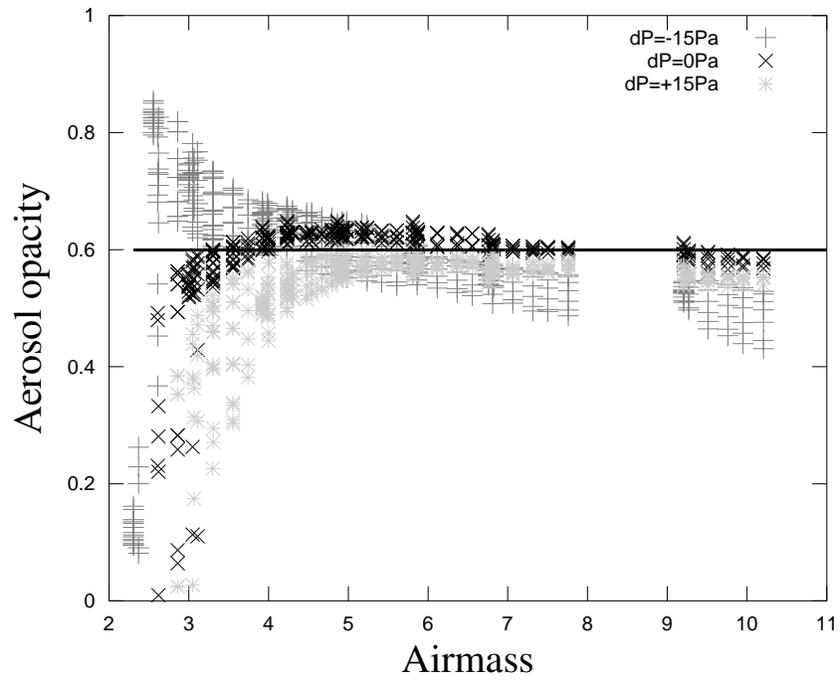}

\caption{\selectlanguage{english}%
Influence of the meteorological variability on the estimation of $\tau_{aer}^{k0}$
depending on airmass $\nu$.\foreignlanguage{french}{\label{fig:sensi_dPsurf}}\selectlanguage{french}%
}
\end{figure}

\begin{figure}
\includegraphics[bb=0bp 150bp 550bp 642bp,clip,width=1.\textwidth]{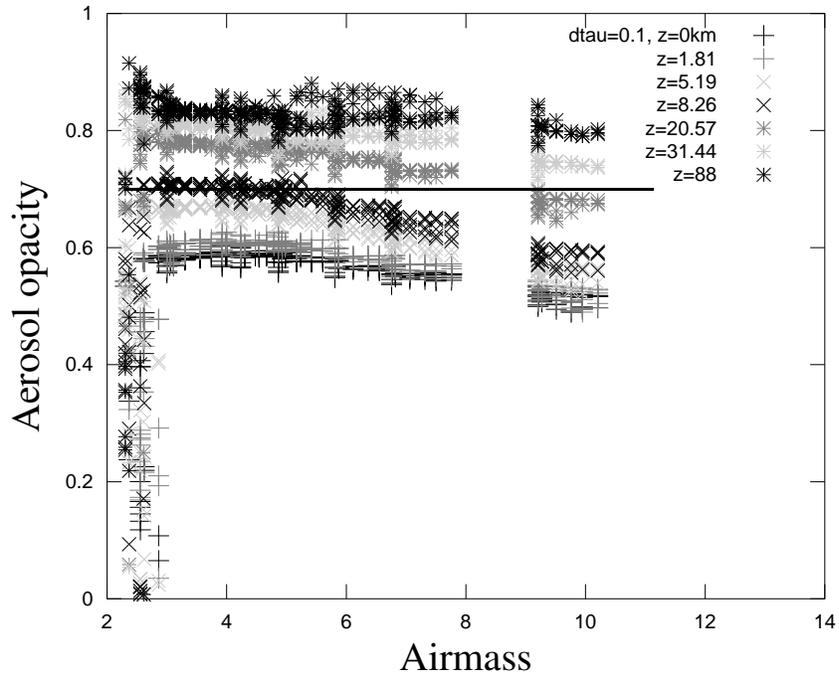}

\caption{\selectlanguage{english}%
Influence on the estimation $\tau_{aer}^{k0}$ of perturbing the reference
vertical distribution profile of the aerosols by a local perturbation
of opacity 0.1 as a function of the airmass $\nu$. \label{fig:sensi_dtau} \selectlanguage{french}%
}
\end{figure}

The validation is realized by inverting synthetic data, that is to
say, realistic TOA spectra $I^{k}$ calculated with DISORT (section
\ref{sub:Couplage-gaz-aerosols}) to which we add zero-mean Gaussian
noise simulated using a meaningful covariance matrix, i.e. calculated
from an estimation of OMEGA noise. In the reference simulation, we
set the properties of the surface $A_{surf}^{k0}=0.3$ (minerals)
and of the atmosphere $\tau_{aer}^{k0}=0.6$, $H_{scale}=11$ km,
and initial pressure $p_{surf}$. In a second simulation, for the
calculation of the spectrum $I^{k}$, we apply a pressure deviation
of $\Delta p_{surf}=\pm15$ Pa typical of the Martian meteorological
variability (barocline activity, \citep{Forget1999}) but the factor
$\beta$ is estimated with a transmission spectrum $T_{gaz}^{k}$
calculated according to the initial pressure. In a third simulation,
we perturb the initial distribution profile of the aerosols (exponential)
by increasing the opacity of one given layer $l,\: l=1,\ldots$ in
succession by an additional amount $\Delta\tau_{aer}^{k0}=0.1$. Thus,
the total AOD becomes 0.7 with the perturbed profile and there are
as many runs as the layer number. In each run, the geometry varies
according to the values given in Table \ref{tab:LUT-beta}. In agreement
with the sensitivity study (section \ref{sub:Sensitivity-and-feasibility}),
Figure \ref{fig:sensi_dPsurf} shows that our method estimates $\tau_{aer}^{k0}$
with an accuracy comparable to the usual methods in the reference
case provided that the airmass exceeds $\nu\approx$ 3. Since we do
not have access to the real ground pressure but to a value predicted
by a MGCM, we expose ourselves to potentially important errors if
$\nu\lesssim4$ but lower than 10\% beyond ($4\lesssim\nu\lesssim8$).
Figure \ref{fig:sensi_dtau} illustrates the fact that a perturbation
of opacity does not have the same impact on factor $\beta$ and thus
on the estimation $\tau_{aer}^{k0}$ according to its height above
the ground. A low lying aerosol layer is not detectable by our method
because it does not have much influence on the mean pathlength of
the photons into the gas. By contrast, a detached layer high in the
atmosphere superimposed on the standard profile has a strong influence
on the coupling. Thus, such a layer could skew our estimation because
the latter does not take into account such a local perturbation. In
conclusion, we state that our method is reliable if two conditions
are full filled: (i) the observation conditions provide an airmass
that is large enough, i.e. $\nu\gtrsim3.5$ (ii) the aerosol particles
are well vertically mixed caused by a vigorous convection.

\selectlanguage{french}%

\selectlanguage{english}%

\section{Results}

\label{sec:results-select-imag}

\subsection{OMEGA}

\begin{figure}
\includegraphics[bb=150bp 80bp 642bp 600bp,clip,width=0.5\paperheight]{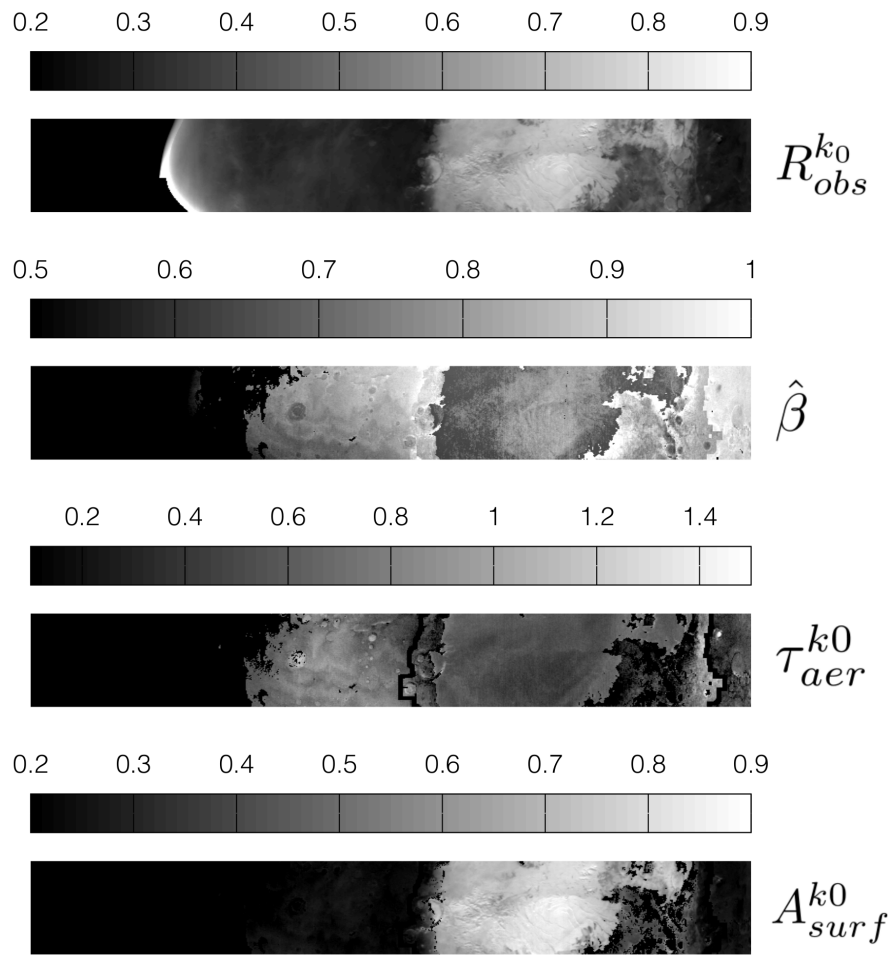}

\caption{Parameter maps for the OMEGA observation \texttt{1880\_1}. From top
to bottom $R_{obs}^{k_{0}}$ , $\hat{\beta}$, $\tau_{aer}^{k0}$
and $A_{surf}^{k_{0}}$.\foreignlanguage{french}{\label{fig:produits_inv_beta}}}
\end{figure}

Figure \ref{fig:produits_inv_beta} shows in a synthetic way all the
products obtained by the chain of operations described above for the
observation OMEGA \texttt{ORB1880\_1} that is representative of many.
The chosen image covers a large geographical area of Mars at high
and medium southern latitudes in spring. The bright extended part
appearing in the maps $R_{obs}^{k_{0}}$ and $A_{surf}^{k_{0}}$ represents
the seasonal deposits of frozen \co\ as well as the permanent cap
that is, at that time, buried by the latter. The black coded area
indicate the surfaces (among then a significant fraction of the ``cryptic''
region - where \co\ very much contaminated by dust exists - for which
no method for estimating $\tau_{aer}^{k0}$ is currently available
or indicate the pixels for which the iterative inversion has not worked
(for example near the limb). 

\begin{figure}
\includegraphics[bb=0bp 150bp 550bp 642bp,clip,width=0.6\paperwidth]{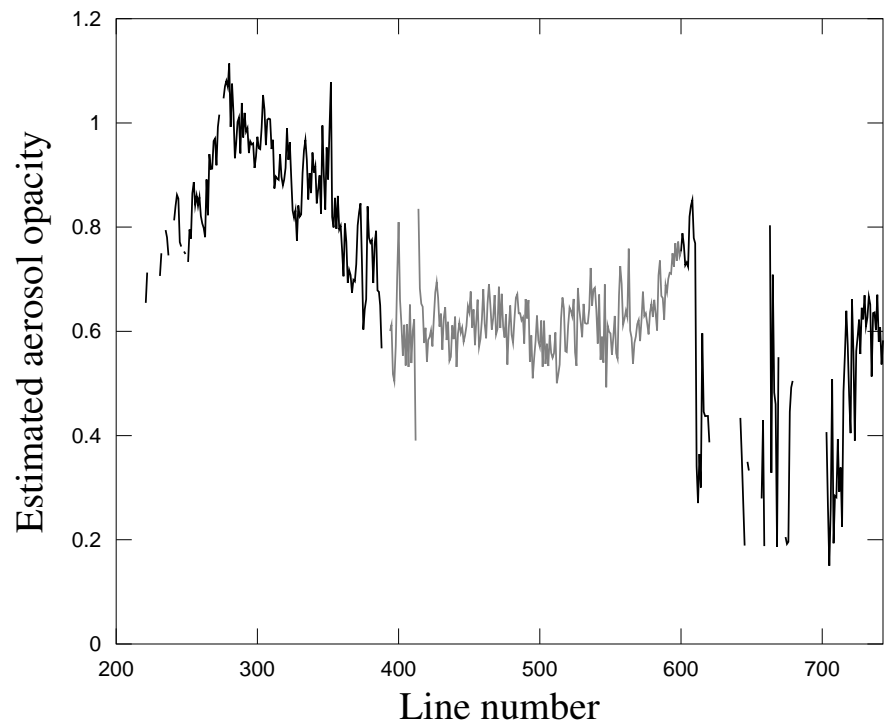}

\caption{A MEX along track-profile of $\tau_{aer}^{k0}$ extracted from the
aerosol optical depth map 1880\_1. The method presented in this paper
(factor $\beta$) produces the black segment while using the method
by \citet{Vincendon2008} leads to the grey segment \label{fig:profiles_1880}}
\end{figure}

First  cross validation of the two methods used in order to produce
the map $\tau_{aer}^{k0}$ - ours and and the one by \citet{Vincendon2008}-
can be performed as follow. We plot along-track profiles of optical
depth through the maps and we examine to which extent the segments
corresponding to the mineral surfaces (in black) and those corresponding
to the frozen surfaces (in grey) are well connected in figure \ref{fig:profiles_1880}.
The right-side red segment with high dispersion corresponds to the
``cryptic'' region. The quality of the connection seems quite good
for \texttt{ORB1880\_1} and for other observations that we treated.

\begin{figure}
\includegraphics[bb=50bp 300bp 600bp 770bp,clip,width=0.7\paperwidth]{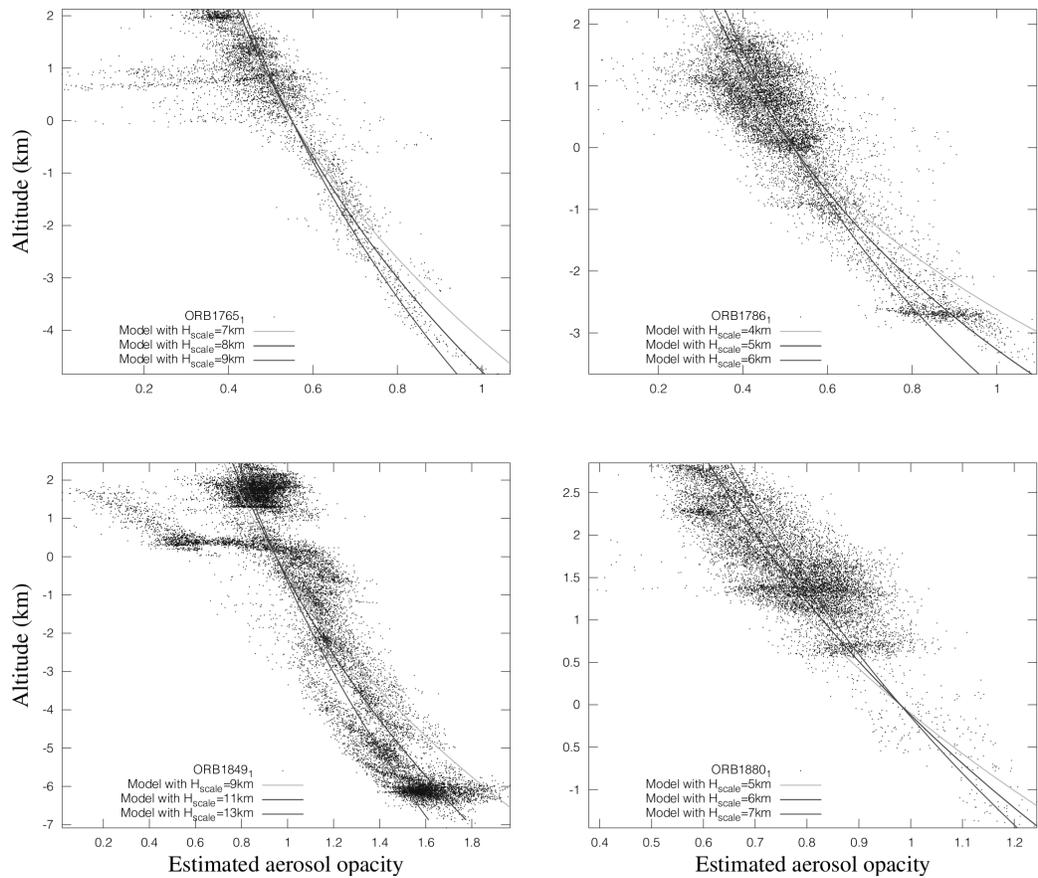}

\caption{Scatter plot of $\tau_{aer}^{k0}$ as a function of altitude $h$
for the OMEGA observations 1765\_1, 1786\_1, 1849\_1, and 1880\_1
( dots). The continuous lines are the result of applying a regression
to the data with an exponential $\tau_{0}\exp\left(-h/H\right)$,
the scale height $H$ varying around the value that insures the best
fit (in black).\label{fig:plot_h_tau_1880}}
\end{figure}

Second the problem of evaluating the dust scale height, which is an
input of our retrieval algorithms, is investigated more thoroughly.
Our region of interest is the southern high latitudes in spring for
which dust activity is monitored in the companion paper. In this case
there are two possibilities for constraining the dust scale height
provided that it can be considered spatially constant for one given
observation. First its value (along with factor $p_{0}$) can be adjusted
by maximizing the overall convergence rate achieved by algorithm 3.5a
when analyzing each image of a selection of OMEGA observations. In
all cases $H_{scale}$=11 km turns out to be the optimal value. Note
that this optimization problem is satisfactorily convex, and thus
its solution satisfactorily constrained, since the convergence of
the iterative inversion is very sensitive to the value of ($p_{0}$,
$H_{scale}$). It is important to keep in mind that the previous scale
height expresses the rate of optical depth change with height. Second
large scale geographical variations of $\tau_{aer}^{k0}$ can be interpreted
as due, at first order, to changes of the atmospheric column linked
with changes of pixel altitude. Indeed if we assume that the intrinsic
optical properties (such as the extinction cross-section $\sigma$)
and the scale height of dust particles are both geographically and
vertically constant and that their density number $N_{0}$ normalized
to altitude 0 is also geographically constant, then $\tau_{aer}^{k0}(h)=N_{0}\sigma H\exp\left(-h/H\right)=\tau_{0}\exp\left(-h/H\right)$.
We are restricted to OMEGA observations that cover an area with a
range of altitudes large enough so that the experimental cloud of
points obtained by plotting $\tau_{aer}^{k0}$ as a function of the
pixel altitude $h$ can be satisfactorily fitted by the model. Figure
\ref{fig:plot_h_tau_1880} shows the result for four global observations.
The data can be satisfactorily explained with values of $H$ varying
from 6 to 11 kilometers, most often in the lower end of this interval,
in agreement with \citet{Vincendon2010b}. We note that the ``spatially''
derived scale height $H$ does not correspond to the ``radiatively''
derived scale height $H_{scale}$ ($H\lesssim H_{scale}$) with the
exception of observation ORB1849\_1. If we trust the method a possible
way to explain this discrepancy is to suppose that $N_{0}$ must show
a overall systematic trend of decrease with altitude. In a sense this
parameter expresses the intensity of dust loading in the atmosphere
regardless of the length of the atmospheric column. We hypothesize
that the systematic trend is linked with the decrease capability of
the atmosphere to lift dust with the decrease of atmospheric density
while the dispersion around the main trend is linked with the meteorological
activity. If we assume also an exponential altitude profile with a
scale height of $H'_{scale}$ for $N_{0}$, the effective ``spatially''
derived scale height for $\tau_{aer}^{k0}$ is $H\approx\nicefrac{H_{scale}H_{'scale}}{H_{scale}+H_{'scale}}$
$\approx$ 6km for $H_{scale}=H'_{scale}$=11 km, consistent with
what we found. Referring to Figure \ref{fig:plot_h_tau_1880}, it
is interesting to note that the observation with the highest spatially
derived scale height corresponds to a peak of dust activity around
$L_{S}\text{\ensuremath{\approx}}230$\textdegree{} also observed
by Pancam during MY27 \citep{Lemmon2006}. This global dust event
erases the overall systematic trend of $N_{0}$ decrease with altitude.

Because it is impossible to constrain $H_{scale}$ for each OMEGA
observation we choose to take a fixed value $H_{scale}$ =11 km. We
have determined that possible excursions of $H_{scale}$ by $\pm$
3 kilometers from the reference value could lead to relative errors
up to 50\%. 

\begin{figure}
\includegraphics[width=0.5\textwidth]{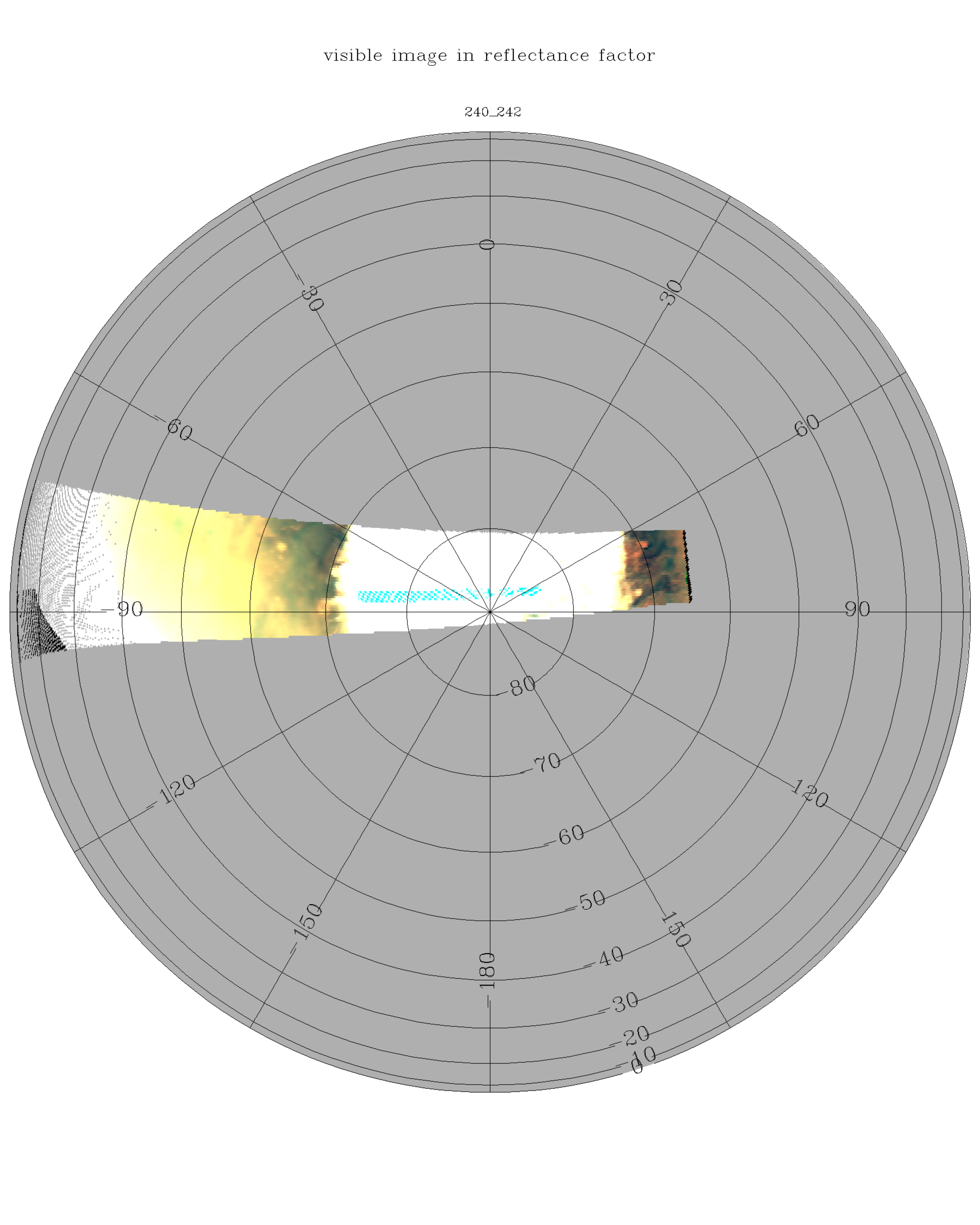}\includegraphics[width=0.5\textwidth]{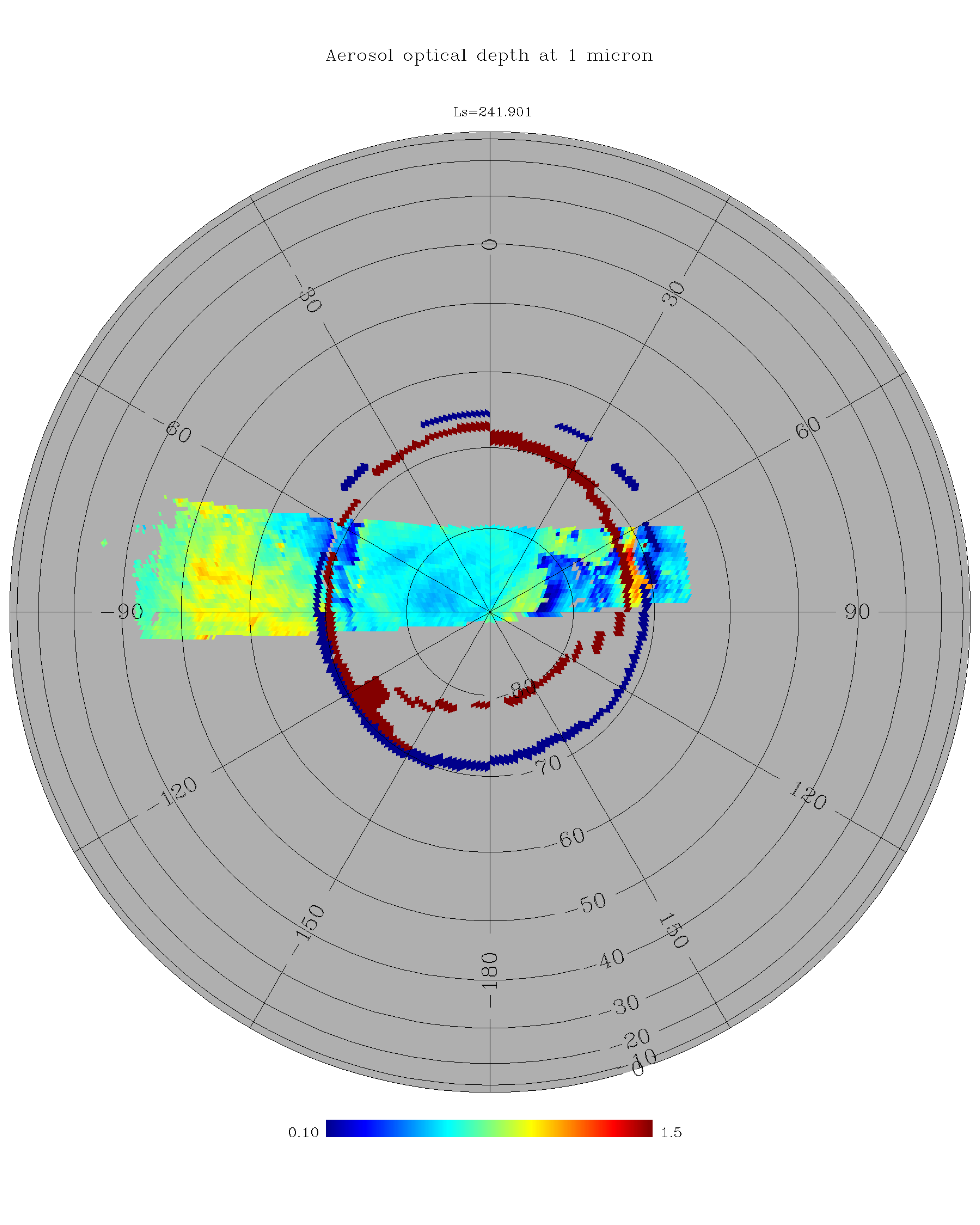}

\caption{\selectlanguage{british}%
Global OMEGA observation \foreignlanguage{english}{\texttt{1880\_1}}.
The  composition on the left displays an RGB composition of the TOA
martian reflectivity in the visible which is stretched so as to reveal
dust in the atmosphere as yellowish hues. The composition on the right
displays the Aerosol Optical Depth map at 1 micron.\label{fig:comp_AOD_map_VIS}\selectlanguage{english}%
}

\end{figure}

Finally an enlightening comparison is put forward between the AOD
retrieved from our analysis of the near infrared \texttt{1880\_1}
image and an RGB composition made by extracting the TOA martian reflectivity
measured by OMEGA at three wavelengths (0.7070, 0.5508, and 0.4760
\micron) from the corresponding visible image (Figure.\ref{fig:comp_AOD_map_VIS}).
The composition is stretched so as to make visible the dust clouds
as yellowish hues against mineral surfaces while the seasonal deposits
appear completed saturated. Excellent qualitative agreement can be
noted between the two products down to the details. In the companion
paper other successful examples are provided. 

\subsection{CRISM}

In the CRISM case, six observations of the landing site of the Mars
Exploration Rover ``Spirit'' (i.e. West of Columbia Hills), acquired
with varied geometry and atmospheric conditions (see Table \ref{tab:resultsCRISM}),
have been chosen to test our method (hereafter identified as the ``$\beta$
method''). For that purpose, these scenes are of special interest
since the AOD values were also retrieved from the same data by Wolff
et. al. (personal communication) and from PanCam synchronous measurements
at the ground looking towards zenith \citep{johnson06b}. The previous
procedures are identified as the ``W method'' and the ``P method''
respectively. Special attention is paid by focusing on areas of the
observations with altitude variations not exceeding one hundred meters
and with negligible slope at the scale of the pixel size and above.
Following the procedure described in section \ref{sec:post-proc-form},
the eleven CRISM images of each FRT observation are combined and binned
at about 300 meters per pixel. Then, the algorithm adapted to CRISM
for the simultaneous retrieval of surface albedo and aerosol opacity
is applied on the CSP cube (section \ref{sec:post-proc-form}). The
operation leads to the results tabulated in Table \ref{tab:resultsCRISM}
along with those obtained by the ``W'' and ``P'' methods. In addition
Figure \ref{fig:beta-curves-FRT-Gusev} illustrates to which extent
our best model matches the experimental $\hat{\beta}$ versus airmass
curves for each observation. Similarly to the OMEGA case, airmass
$\nu\approx3$ represents a threshold above which the fit becomes
very satisfactory with the exception of the observation FRT95B8. One
should also note that the larger the difference between the two azimuthal
modes is, the more constrained the best solution likely becomes because
the two branches of the $\hat{\beta}$ versus airmass curves are increasingly
separated.

Once the content of aerosols is known, CRISM TOA radiances can be
corrected in order to retrieve surface BRF. The correction of remotely
sensed images for atmospheric effects is however not straightforward
due to the anisotropic scattering properties of the atmospheric aerosols
and the materials at the surface. Traditional inversion algorithms
are based on reductionist hypothesis that assumes that the surface
is lambertian \citep{McGuire2008}. Although this assumption largely
simplifies the inverse problem, it critically corrupts the angular
shape of the retrieved BRF since solid surfaces are hardly isotropic
\citep{lyapustin99a}. Recently, we have proposed an original inversion
method to overcome these limitations when treating CRISM multi-angle
observations \citet{Ceamanos2013}. This inversion algorithm is based
on a TOA radiance model that depends on the Green\textquoteright{}s
function of the atmosphere and a semi-analytical expression of the
surface BRF. In this way, robust and fast inversions of the model
on CRISM TOA radiance curves are performed accounting for the anisotropy
of the aerosols and the surface. The root mean square error achieved
by the model when reproducing the CRISM TOA reflectance factor curves
is indicated in Table \ref{tab:resultsCRISM} for each observation
and for each AOD input. A NaN value indicates that the inversion was
not successful, i.e. no valid surface BRF could be retrieved with
the proposed aerosol properties and atmospheric opacity. 

By examining Table \ref{tab:resultsCRISM} we conclude that our method
gives results generally in good agreement with the values retrieved
by Wolff et al. with the notable exception of observations 812F and
95B8. In the first case, the ``W'' method overestimates $\tau_{aer}^{k0}$
since, with such an opacity, the inversion algorithm has difficulties
to produce a physically meaningful surface reflectance. At contrary
our value is in agreement with the one estimated from the PanCam measurements.
In the second case, the $\hat{\beta}$ versus airmass curves are not
well fitted by our model implying that the estimated value for $\tau_{aer}^{k0}$
by the ``$\beta$ method'' may not be accurate. One should also
note that the root mean square error achieved by the TOA reflectance
factor modeling using our value is systematically lower than when
using the value given by the ``W method'' except for observation
334D. We put forward the hypothesis that the AOD estimate by the ``$\beta$
method'' is less biased even though the latter also uses the lambertian
surface hypothesis in the radiative transfer calculations. As regards
to the previous hypothesis, the relatively weak dependence of the
main parameter $\hat{\beta}$ from which the AOD is retrieved is a
possible explanation. Indeed the radiative coupling between aerosols
and gas is mostly expressed in the atmospheric additive term and,
to a lesser extent, the terms including multiple scattering of photons
between surface and atmosphere. In the first case at least the anisotropy
of the surface has no influence.\textbf{ }By contrast, methods that
directly invert a lambertian surface-atmosphere RT model
on the TOA radiance in the continuum of the spectra, such as the ``W
method'' lead to an AOD estimate that is more sensitive to the surface
bidirectionnal reflectance.  We explain the moderate discrepancies
of results that can be observed between the ``P method'' and the
``$\beta$ method'' if we consider that the first one is more sensitive
to the low lying layers of aerosols whereas the second one is more
sensitive to aerosols layers situated at $\approx$ 2 km in height
and above (see section \ref{sub:Validation}).

\section*{Conclusions}

\label{sec:conclusion}

In this article we propose a method to retrieve the optical depth
$\tau_{aer}^{k0}$ of Martian aerosols from OMEGA and CRISM hyperspectral
imagery. The method is based on parametrization of the radiative coupling
between particles and gas determining, with local altimetry, acquisition
geometry, and the meteorological situation, the absorption band depth
of gaseous \co. The coupling depends on (i) the acquisition geometry
(ii) the type, abundance and vertical distribution of particles, and
(iii) the surface albedo $A_{surf}^{k}$. For each spectrum of an
image, we compare the depth of the 2 \micron\ absorption band of
gaseous \co\  between (i) the observed spectrum and (ii) the corresponding
transmission spectrum through the atmospheric gases alone. The latter
is calculated with a line-by-line RT model fed by the vertical compositional
and thermal profiles predicted by the European Mars Climate Database
for the dates, locations, and altitudes of the observations. Thus
we define a significant new parameter $\beta$ that expresses the
strength of the gas-aerosols coupling while directly depending on
$\tau_{aer}^{k0}$. Combining $\beta$ and the radiance value $R_{obs}^{k_{1}}$
within the 2 \micron\ band, we evaluate $\tau_{aer}^{k0}$ and $A_{surf}^{k}$
by radiative transfer inversion and provided that the radiative properties
of the aerosols are known from previous studies and that an independent
estimation of the scale height of the aerosols is available. One should
note that practically $\beta$  can be estimated for a large variety
of mineral or icy surfaces with the exception of CO\textsubscript{2}
ice when its 2 microns solid band is not sufficiently saturated.

The validation of the method was performed both with synthetic and
real data. It shows that our method is reliable if two conditions
are fulfilled: (i) the observation conditions provide  large incidence
or/and emergence angles (ii) the aerosol are vertically well mixed
in the atmosphere. A low lying aerosol layer is not detectable by
our method because it does not have much influence on the mean pathlength
of the photons into the gas. By contrast, a detached layer high in
the atmosphere superimposed on the standard profile has a strong influence
on the coupling. Our method works even if the underlying surface is
completely made of minerals, corresponding to a low contrast between
surface and atmospheric dust, while being observed at a fixed geometry.
This is the first principal asset of our method. Minimizing the effect
of the surface Lambertian assumption on the AOD retrieval is the second
principal asset of our method.

Experiments conducted on OMEGA nadir looking observations -as well
as CRISM EPF- produce very satisfactory results. With OMEGA, we note
a good coherency between our approach and the one of \citet{Vincendon2008}.
 The domain of airmass ($\nu\gtrsim3.5$) for which our method is
reliable implies that it is intended for analyzing high latitude OMEGA
observations with sufficiently high solar zenith angles ($\gtrsim$65\textdegree{}).
This constraint is somewhat lighten with CRISM EPF observations that
imply a large range of emergence angles and thus values of airmass.
Indeed we note very satisfactory $\tau_{aer}^{k0}$ retrievals for
solar incidence angle down to 33\textdegree{} extending the applicability
of the method to non polar regions. Finally we should note that our
method was applied for the first time extensively on a series of OMEGA
observations in order to map the atmospheric dust opacity at high
southern latitudes from early to late spring of Martian Year 27. This
study is presented in the companion paper.

\section*{Acknowledgments}

This work was done within the framework of the Vahiné project funded
by the ``Agence Nationale de la Recherche'' (ANR) and the ``Centre
d'Etudes Spatiales'' (CNES). 

\begin{flushleft}
{\scriptsize \bibliographystyle{elsarticle-harv}

} 
\par\end{flushleft}

\newpage

\begin{table}
\begin{tabular}{|c|c|c|}
\hline 
 & Gas & Aerosols\tabularnewline
\hline 
\hline 
Single scattering albedo & $\omega_{gas}=0$ & $\omega_{aer}$\tabularnewline
\hline 
Optical depth & $\tau_{gas\, i}\:,\Delta g_{i}$, $i=0,\ldots,N+1$ & $\tau_{aer}$\tabularnewline
\hline 
Phase function & $\Upsilon_{gas}=0$ & $\Upsilon_{aer}$\tabularnewline
\hline 
\multicolumn{3}{|c|}{$\tau_{i}=\tau_{gas\, i}+\tau_{aer}$ , $\omega_{i}=\omega_{aer}/(1+\frac{\tau_{gas\, i}}{\tau_{aer}})$,
$i=0,\ldots,N+1$ $\Upsilon=\Upsilon_{aer}$}\tabularnewline
\hline 
\end{tabular}

\caption{Sum up of the physical quantities characterizing the ``elementary''
processes of the radiative transfer taking place in each homogeneous
atmospheric layer. We omit in the notation the indices $l$ and $k$
for readability.\foreignlanguage{french}{\label{tab:quant_rad_atmos}}}
\end{table}

\clearpage

\begin{table}
\begin{tabular}{|c|c|c|c|c|c|}
\hline 
 & $p_{0}$ & $p_{1}$ & $p_{2}$ & $p_{3}$ & $p_{4}$\tabularnewline
\hline 
\hline 
OMEGA & 1.2 & 0.66312127  & 0.44429186 & -0.024559039 & 0.00068174262\tabularnewline
\hline 
CRISM & 0.95 & 0.66312127  & 0.44429186 & -0.024559039 & 0.00068174262\tabularnewline
\hline 
\end{tabular}

\caption{Value of the coefficients allowing to calculate $\psi(\nu)$.\label{tab:coef_psi} }
\end{table}

\clearpage

\begin{table}
\begin{tabular}{|c|c|c|c|c|c|c|c|c|c|c|c|c|c|}
\hline 
$\theta_{i}$ (\textdegree{}) & 0  & 20 & 40 & 50 & 60 & 70 & 72 & 75 & 78 & 80 & 83 &  & \tabularnewline
\hline 
\hline 
$\theta_{e}$ (\textdegree{}) & 0 & 2  & 20 & 40 & 50 & 55 & 60 & 70 &  &  &  &  & \tabularnewline
\hline 
$\phi_{e}$ (\textdegree{}) & 0  & 30 & 60 & 90 & 120 & 150 & 180 &  &  &  &  &  & \tabularnewline
\hline 
$\tau_{aer}^{k0}$ & 0.005 & 0.01 & 0.05 & 0.1 & 0.2 & 0.3 & 0.4 & 0.5 & 0.6 & 0.7 & 0.8 & 0.9 & 1.0\tabularnewline
\hline 
 & 1.2 & 1.4 & 1.6 & 1.8 & 2. & 2.2 & 2.4 & 2.6 & 2.8 & 3.  & 3.5 & 4.0 & 4.5\tabularnewline
\hline 
$A_{surf}^{k_{1}}$\footnotemark[1] & 0 & 0.01 & 0.05 & 0.1 & 0.2 & 0.3 & 0.4 & 0.5 & 0.6 &  &  &  & \tabularnewline
\hline 
$H_{scale}$ (km) & 4 & 6 & 8 & 11 & 14 & 17 & 20 &  &  &  &  &  & \tabularnewline
\hline 
\end{tabular}

\caption{Series of values taken by the incident, emergent and azimuth angles
as well as by the physical parameters used for the construction of
the reference LUT\label{tab:LUT-beta}}
\end{table}

\clearpage

\begin{table}
\begin{tabular}{|c|c|c|c|c|c|c|}
\hline 
Observation FRT & 334D & 7D6C & 812F & 95B8 & B6B5 & 3192\tabularnewline
\hline 
Incidence angle & 55.4 & 36.4  & 32.6 & 39.3  & 56.4 & 60.4 \tabularnewline
\hline 
Azimutal modes & 64, 118\textdegree{} & 68, 104\textdegree{} & 88, 97\textdegree{} & 71, 113\textdegree{} & 51, 129\textdegree{} & 65, 135\textdegree{}\tabularnewline
\hline 
 &  &  &  &  &  & \tabularnewline
\hline 
$\tau_{aer}^{k0}$ ``W method'' (1 \micron) & 0.35 & 1.74 & 1.92 & 0.55 & 0.35 & 0.33\tabularnewline
\hline 
RMSE w/model & 0.877E-02 & 2.39E-02 & 1.58E-02 & 1.12E-02 & 2.47E-02 & 2.21E-02\tabularnewline
\hline 
 &  &  &  &  &  & \tabularnewline
\hline 
$\tau_{aer}^{k0}$ ``P method'' (1 \micron) & 0.871 & 1.32 & 0.93 & 0.66 & 0.48 & 0.31\tabularnewline
\hline 
RMSE w/model & NaN & 2.02E-02 & 1.05E-02 & 1.42E-02 & 1.55E-02 & 2.38E-02\tabularnewline
\hline 
 &  &  &  &  &  & \tabularnewline
\hline 
$\tau_{aer}^{k0}$ ``$\text{\ensuremath{\beta}}$ method'' (1 \micron) & 0.46 & 1.41 & 1.00 & 0.36 & 0.46 & 0.38\tabularnewline
\hline 
RMSE w/model & 1.10E-02 & 2.13E-02 & 1.10E-02 & 0.877E-02 & 1.65E-02 & 1.56E-02\tabularnewline
\hline 
\end{tabular}

\caption{Optical depth at 1 \micron\ retrieved by three different methods
for our selection of observations. The Root Mean Square Error (RMSE)
refers to the adequacy of a non Lambertian surface-atmosphere radiative
transfer model fed by the latter optical depth in reproducing the
TOA reflectance factors measured by CRISM. \label{tab:resultsCRISM}}
\end{table}

\clearpage

\end{document}